\documentclass{emulateapj}
\usepackage{natbib}

\newcommand\etal{et al.}
\newcommand\ie{i.e.}
\newcommand\eg{e.g.}
\newcommand\msun{{\rm M_\odot}}
\newcommand\kms{\ifmmode{\rm km\ s^{-1}}\else$\rm km\ s^{-1}$\fi}
\def\eps@scaling{1.0}
\newcommand\plotrtwo[2]{{
 \typeout{Plotrtwo included the files #1 #2}
 \centering
 \leavevmode
 \columnwidth=.45\columnwidth
 \includegraphics[angle=270,width={\eps@scaling\columnwidth}]{#1}
 \hfil
 \includegraphics[angle=270,width={\eps@scaling\columnwidth}]{#2}
}}
\newcommand\plotthree[3]{{
 \typeout{Plotthree included the files #1 #2 #3}
 \centering
 \leavevmode
 \columnwidth=.30\columnwidth
 \includegraphics[width={\eps@scaling\columnwidth}]{#1}
 \includegraphics[width={\eps@scaling\columnwidth}]{#2}
 \includegraphics[width={\eps@scaling\columnwidth}]{#3}
}}

\shorttitle{N5098 FROM CHANDRA}
\shortauthors{RANDALL ET AL.}
\slugcomment{Astrophysical Journal, accepted}

\begin{document}

\title{Gas Sloshing and Bubbles in the Galaxy Group NGC~5098}

\author{S.\ W.\ Randall\altaffilmark{1}, C.\ Jones\altaffilmark{1},
  M.\ Markevitch\altaffilmark{1},  E.\
  L.\ Blanton\altaffilmark{2}, P.\ E.\ J.\ Nulsen\altaffilmark{1}, W.\
  R.\ Forman\altaffilmark{1}}

\altaffiltext{1}{Harvard-Smithsonian Center for Astrophysics, 60
  Garden St., Cambridge, MA 02138, USA; srandall@cfa.harvard.edu,
  cjf@cfa.harvard.edu, mmarkevitch@cfa.harvard.edu,
  wrf@cfa.harvard.edu}
\altaffiltext{2}{Institute for Astrophysical Research, Boston
  University, 725 Commonwealth Ave., Boston, MA 02215, USA; eblanton@bu.edu}

\begin{abstract}
We present results from {\it Chandra} observations of the galaxy pair
and associated galaxy group NGC~5098, and find evidence for both gas
sloshing and AGN heating.  The X-ray brightness images
show diffuse emission with a spiral structure, centered on NGC~5098a,
and a sharp edge in the 
diffuse emission surrounding much of the galaxy at about 30~kpc.  The
spiral structure in the X-ray surface brightness and temperature maps,
the offset between the peak of the cool gas and the central AGN, and
the structure of the cold front edges all suggest gas sloshing in the core.  The
most likely perturber is the nearby galaxy NGC~5098b, which has been
stripped of its gaseous atmosphere.  Detailed images of the core
reveal several X-ray cavities, two of which, just north and southeast
of the central AGN, correlate with radio emission and have bright
X-ray rims, similar to buoyant bubbles seen in the ICM of other
systems.  We estimate the pressures in the
bubbles and rims and show that they are roughly equal, consistent with
these being young features, as suggested by their close proximity to
the central AGN.  We assume that
the other X-ray cavities in the core, which show no correlation with
existing radio observations, are ghost cavities from previous AGN
outbursts.  An estimate of the mechanical energy required to inflate
the cavities indicates that it is sufficient to offset radiative
cooling of the gas for 15~Myr.  Therefore, for a typical cycle time of
$10^7$~yrs, the central AGN energy output is enough to balance cooling
over long timescales.
\end{abstract}


\section{Introduction} \label{sec:intro}

A major surprise from early {\it Chandra} and {\it XMM-Newton}
observations was that gas in cool core clusters does not reach the low
central temperatures predicted by radiative cooling models, in
disagreement with the previously accepted cooling flow model (Peterson
\& Fabian 2006). The implication is that the central gas must
experience some kind of heating.  The
source of this heating, and understanding when and how it takes place,
has recently been a major topic of study in extragalactic
astrophysics. A promising candidate is feedback from energy injection
by the central AGN of the cD galaxy (McNamara \& Nulsen 2007).  
However, the details of this interaction, and how the energy is
transferred from the jets to the ambient ICM, are poorly understood.
Galaxy groups provide an excellent opportunity to study heating and
other non-gravitational processes in the ICM.  Although not as X-ray
luminous as clusters, the effects of heating are more readily seen in
groups, due to their lower mass and central density.  For example, the
gas fraction in groups shows a relatively large scatter ($\sim2$ 
at any fixed temperature) within $r_{2500}$, with the scatter being
tightly correlated with the central entropy (Gastaldello \etal\ 2007;
Sun \etal\ 2009),
reflecting the greater role of non-gravitational 
processes in the centers of groups as compared to clusters.

Another discovery from {\it Chandra} was the existence of contact
discontinuities, or cold fronts, where a cool, dense subclump of gas
exhibits a temperature and density jump at the interface with warmer
gas, such that the pressure profile across the interface is
continuous (Markevitch \& Vikhlinin 2007).  More recently, it has been
shown that such cold fronts can be generated not only from subgroup cores
in a merger, but also by gas sloshing around a potential minimum,
caused by an off-axis interaction with a perturber (Ascasibar \&
Markevitch 2006).  Cold fronts are found in clusters and groups with
relatively high frequency, and sloshing cold fronts have been identified in a
handful of systems (\eg, Mazzotta \etal\ 2001; Dupke \etal\ 2007;
Gastaldello \etal\ 2009).
Understanding cold fronts and sloshing is of interest as they can have
a significant impact on cluster cores through gas heating, ICM mixing
and enrichment, turbulence, constraints on conduction and magnetic
fields, etc.

In this paper we report on {\it Chandra} observations of the NGC~5098
galaxy group (RGH~80), originally identified by Ramella \etal\ (1989),
which shows evidence for {\it both} AGN heating and gas sloshing.
Studies of {\it ROSAT} and {\it ASCA} observations found average
temperatures and metallicities for this system of $\sim1$~keV
and $\sim30\%$~solar (Davis \etal\ 1999; Hwang \etal\ 1999; Buote
2000; Mahdavi \etal\ 2000).  More recently, {\it XMM-Newton} observations
were used to derive radial profiles for various
properties of the X-ray gas,
including temperature, pressure, entropy, total mass, gas mass, and
cooling time (Xue \etal\ 2004; Mahdavi \etal\ 2005).  A joint analysis
of the {\it XMM-Newton} and {\it Chandra} data we consider here was
performed by Gastaldello \etal\ (2007) as part of a sample of relaxed
galaxy groups.  The {\it Chandra} data were used to study the global
properties of the gas in this system as part of the group sample
studied by Sun \etal\ (2009).
X-ray observations imply a total group mass of $4-6
\times 10^{13} \msun$, and optical studies find a group line-of-sight
velocity dispersion of $\sigma_{\rm los} = 602$~\kms\ (Xue \etal\ 2004;
Mahdavi \etal\ 2005).

The NGC~5098 galaxy group is dominated by the central galaxy pair
NGC~5098a and NGC~5098b (first identified by Ramella \etal\ 1995,
their ``group 80'').
NGC~5098a is the brighter (and presumably larger) of the two, with
absolute optical magnitudes of $M_B = -21.131$ and $M_V = -22.097$, as compared
to $M_B = -20.845$ and $M_V = -21.770$ for NGC~5098b.  Absolute
magnitudes were calculated from magnitudes given in the Sloan Digital
Sky Survey (SDSS; Adelman-McCarthy \etal\ 2008) catalog and
transformed to the Johnson filter system 
using the relations provided by Smith \etal\ (2002).
The relative line-of-sight velocity of the pair is 360~\kms, slightly
less than the group velocity dispersion.  The western galaxy,
NGC~5098a, hosts the extended radio source B2~1317+33, which has been
detected at several frequencies (Colla \etal\ 1970; Parma \etal\ 1986;
Morganti \etal\ 1997; Condon \etal\ 1998).

We report here on {\it Chandra} observations of NGC~5098.  As the
global structure of the gas has already been studied using
{\it XMM-Newton} (Xue \etal\ 2004) and {\it Chandra} (Sun \etal\ 2009), we 
focus on detailed structure in the central regions, a task for which
{\it Chandra} is well-suited.
The observations and data reduction techniques are described in
\S~\ref{sec:obs}. The X-ray image is presented in \S~\ref{sec:img},
and results on temperature and abundance structure from spectral
analysis are given in \S~\ref{sec:spec}. In \S~\ref{sec:discuss}, we
argue that the central gas is currently sloshing due to a recent interaction
with a nearby galaxy, most likely with NGC~5098b.  We also report on
the detection of several X-ray cavities in the central region and use
these to show that the energy output by the central AGN is sufficient
to balance radiative cooling of the gas. Our results are summarized in
\S~\ref{sec:summary}. 

We assume an angular diameter distance to NGC~5098 of 153.1~Mpc, which
gives a scale of 
0.74~kpc/\arcsec\ for $\Omega_0 = 0.3$,  
$\Omega_{\Lambda} = 0.7$, and $H_0 = 70$~\kms~Mpc$^{-1}$.  All
error ranges are 68\% confidence intervals (\ie, 1-$\sigma$), unless
otherwise stated.

\section{Observations and Data Reduction} \label{sec:obs}

NGC~5098 was originally observed with {\it Chandra} on August 4, 2001,
for 11~ksec with the {\it Chandra} CCD Imaging Spectrometer
(ACIS) in Very Faint mode,  pointed such that the galaxy was visible
on the front-side 
illuminated ACIS-I CCD array.
It was again observed on November 5, 2005, for
39~ksec with ACIS in Very Faint mode, pointed such that the galaxy was visible
on the back-side 
illuminated ACIS-S3 CCD.  
Due to the longer exposure time and better sensitivity at soft
energies, we considered only data from the more recent ACIS-S3 observation.
These data were reduced using the method
described in Randall \etal\ (2008). All data were reprocessed from the
level 1 event files using the latest 
calibration files (as of {\sc CIAO4.0}).  CTI and time-dependent
gain corrections were applied where applicable. {\sc LC\_CLEAN} was
used to remove background
flares\footnote{\url{http://asc.harvard.edu/contrib/maxim/acisbg/}}.
The mean event rate was calculated using time bins within 3$\sigma$ of the
overall mean, and bins outside a factor of 1.2 of this mean were
discarded. The resulting cleaned exposure time was 38.4~ksec.

The emission from NGC5098 and the surrounding group fills the
ACIS-S3 image field of view.  We therefore used the
standard {\sc CALDB\footnote{\url{http://cxc.harvard.edu/caldb/}}}
blank sky background files appropriate for each observation,
normalized to our observations from the ACIS-S1 chip in the 10-12 keV energy band.  To generate exposure maps, we assumed a MEKAL model with $kT = 1$~keV, Galactic absorption, and abundance of 30\%
solar at a redshift $z = 0.0379$, which is consistent with typical results from
detailed spectral fits (see \S~\ref{sec:spec}).

\section{The X-ray Image} \label{sec:img}

\begin{figure}
\plottwo{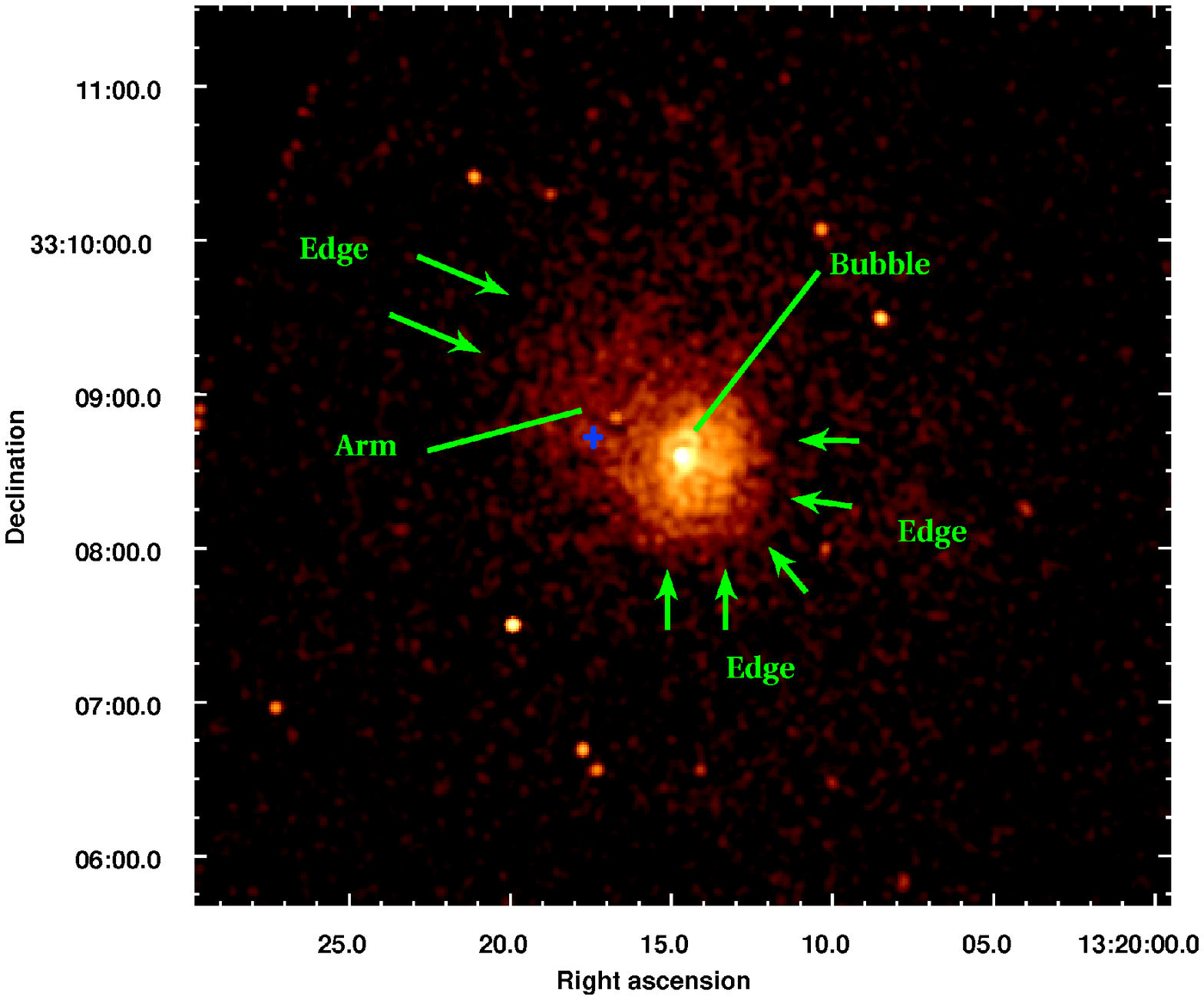}{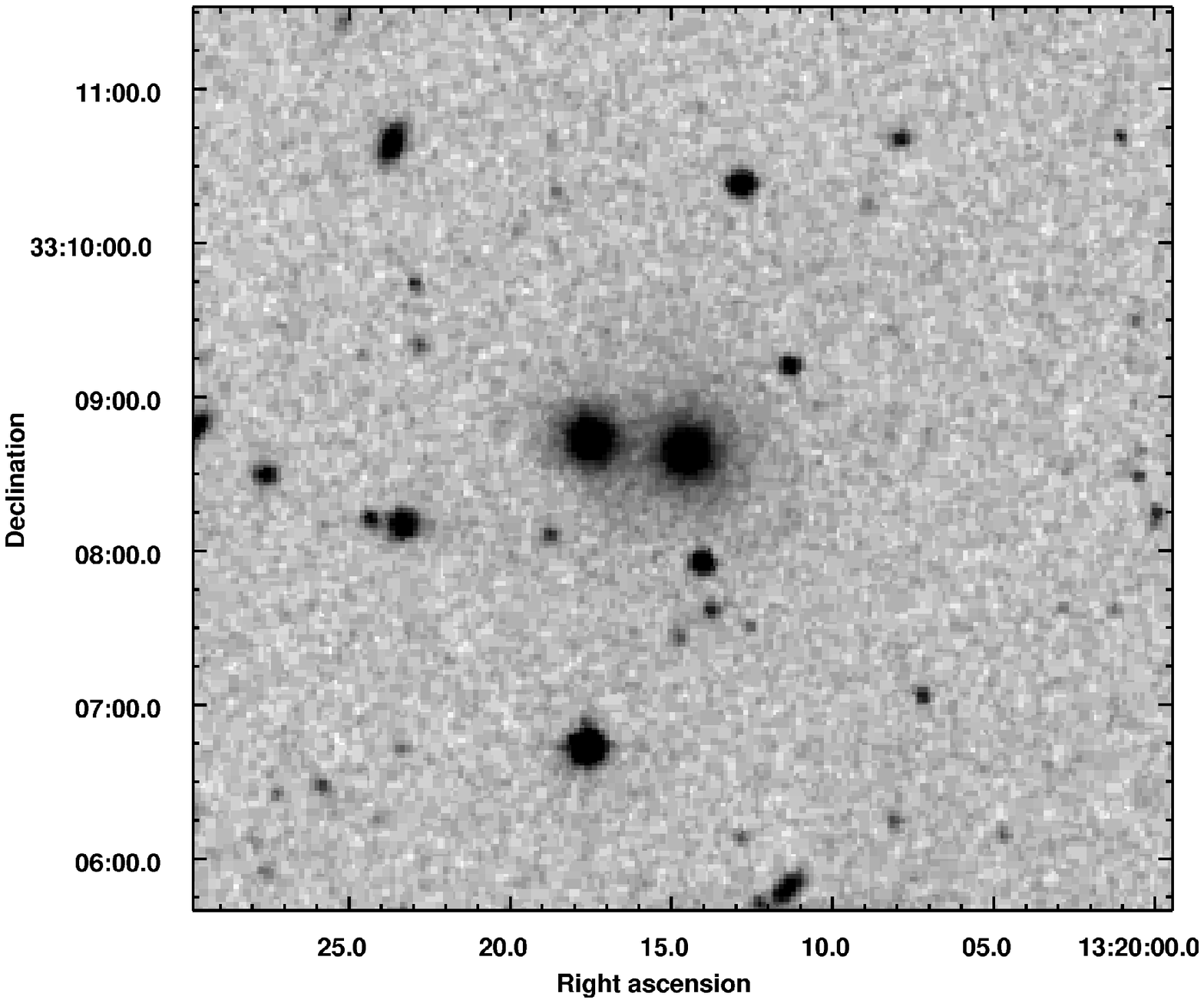}
\caption{
  {\it Left Panel:} Exposure corrected, background subtracted 0.6--5
  keV {\it Chandra} ACIS-S3 observation of 
  NGC~5098.  The image has
  been smoothed with an 3\arcsec\ radius gaussian.   Regions with less than 
  10\% of the total exposure were omitted.
  The image shows a spiral arm structure to the northeast, a sharp surface
  brightness edge beginning in the west and connecting around the
  south to the
  outer boundary of the arm, and central bubbles surrounded by bright
  rims of emission.  The blue cross marks the optical position of NGC~5098b.
  {\it Right Panel:} {\it DSS} image of the same field.  NGC~5098a is
  the western galaxy of the central bright galaxy pair while NGC~5098b
  is to the east.
  \label{fig:fullimg}
}
\end{figure}

The exposure corrected, background
subtracted, smoothed image is shown in
Figure~\ref{fig:fullimg} (the optical {\it DSS} image of the same
field is shown for comparison).  The image shows several interesting
features associated with NGC~5098a:

\begin{itemize}

\item
A bright point source, coincident with the central AGN, near the
center of the diffuse emission.

\item
A plume of emission extending to the northeast
(the ``tail'' noted previously by Gastaldello \etal\ 2007).  The plume
exhibits a 
spiral arm morphology, originating east of NGC~5098a and wrapping
around to the north.  The presence of this feature indicates that the
system is not dynamically relaxed.  NGC~5098a is most likely currently
interacting with NGC~5098b.

\item
A sharp surface brightness edge to the west, southwest, and south,
roughly 30\arcsec\ (28~kpc) from the central AGN.  The edge appears to
continue to the east and define the outer boundary of the arm, adding
to the overall impression of a spiral pattern in the diffuse emission.
The edges are similar to features seen from cold fronts generated by
gas sloshing 
in observations (Dupke \etal\ 2007; Gastaldello \etal\ 2009) and
simulations (Ascasibar \& Markevitch 2006) of 
galaxy clusters and groups.

\item
An asymmetry in the brightest (central) diffuse emission, which extends
farther west of the AGN than to the east, also suggesting that this
system is disturbed.

\item
Two small cavities, surrounded by bright rims of emission,
roughly 4\arcsec\ (3~kpc) to the north and southwest of the AGN.  They
are morphologically similar to ``bubbles'' seen in X-ray observations
of other galaxies and clusters, some of which are associated with
radio emission
(\eg, Abell~2052, 
Blanton \etal\ 2003; Perseus 
cluster, Fabian \etal\ 2006; NGC~4552, Machacek \etal\ 2006;
MS0735.6+7421, McNamara \etal\ 2007; M84,
Finoguenov \etal\ 2008; NGC~5044, Gastaldello \etal\ 2009), though they
are relatively small in
size. The proximity of the bubbles to the AGN suggests that they are
young features, possibly currently being inflated by jets from the AGN.

\end{itemize}

Also of note is the complete lack of X-ray emission associated with
NGC~5098b.  Although the galaxy is well within the northeastern arm of
NGC~5098a, there is no indication of enhanced surface brightness
centered on the galaxy.  This suggests that NGC~5098b may have been
stripped of its gas during an interaction with NGC~5098a or the ambient
group ICM.

\begin{figure}
\plotone{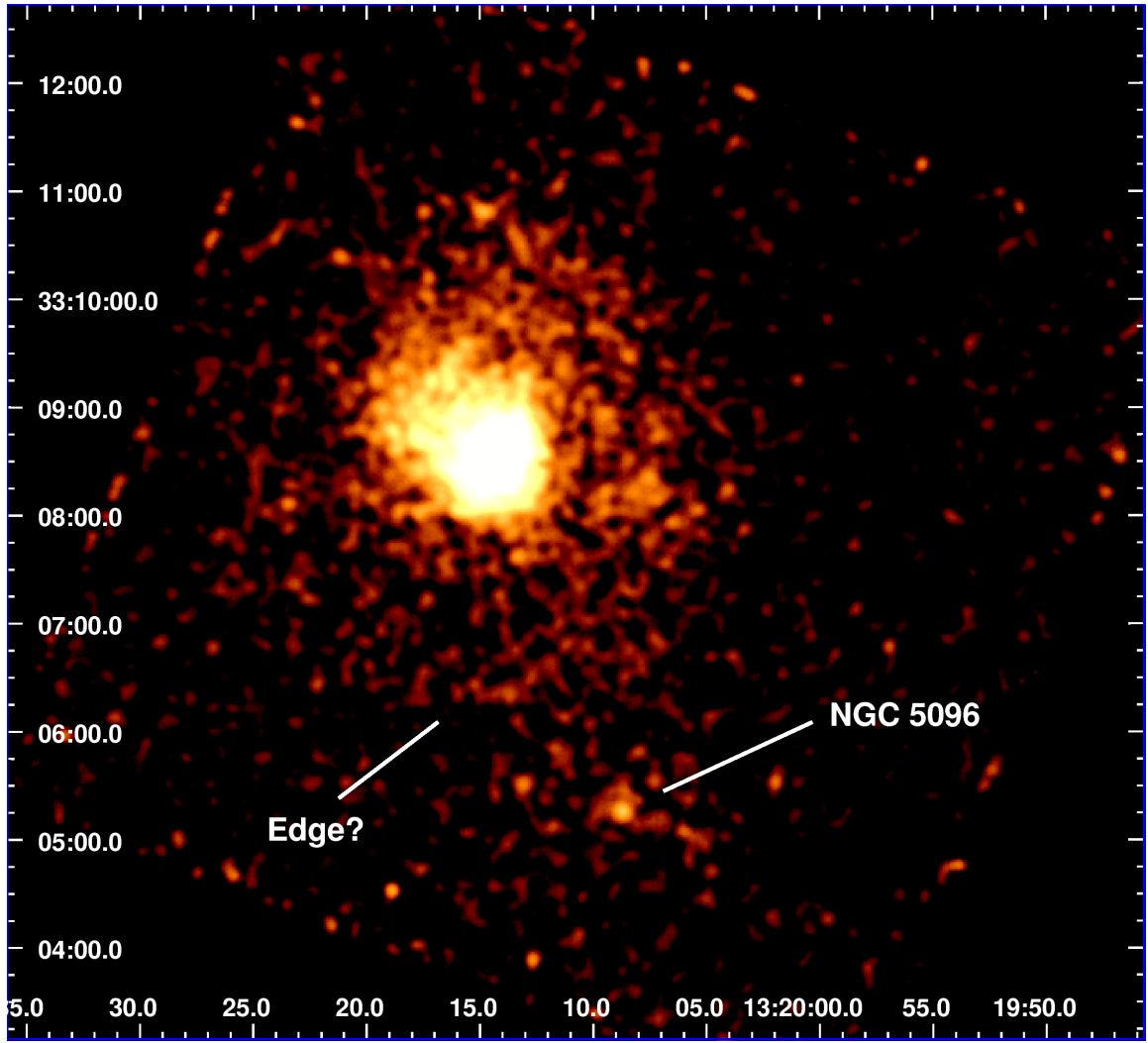}
\caption{
The 0.6--5.0~keV {\it Chandra} image, smoothed with a 12\arcsec\
radius gaussian to better show faint, diffuse emission at large
radii.  Point sources have been removed, as described in the text (see
\S~\ref{sec:img}).  There is a sharp, linear edge in the emission to
the south.  Diffuse emission from the group member NGC~5096 is seen to
the southwest. 
\label{fig:smoothed}
}
\end{figure}

In order to examine the faint, diffuse, group emission at larger radii,
we made a more heavily smoothed image, shown in Figure~\ref{fig:smoothed}. Point sources have
been removed by filling in source regions using a Poisson distribution
whose mean was equal to that of a local annular background region.  
Diffuse emission is seen beyond the core and the edges noted
in Figure~\ref{fig:fullimg}, in all directions.  The extended halo also
shows hints of a spiral structure, continuing from the inner arm noted
in Figure~\ref{fig:fullimg} and wrapping around from the north out to a radius of
$\sim120$\arcsec\ (89~kpc), through the west, and back to the south
out to $\sim132$\arcsec (98~kpc), though the reality of this structure
is unclear, and may just be an artifact of the brighter, inner arm
reaching into the extended emission in the north.
There appears to be a sharp, linear edge in the extended emission to
the south.  The orientation of the edge is such that it cannot be due
to a chip node boundary.  We searched the NASA/IPAC Extragalactic
Database for a foreground absorber that could be responsible for this
feature, but found none.
To the southwest, there is an extended source associated with the galaxy
triplet NGC~5096, which has been included as a member of the NGC~5098
group (Ramella \etal\ 1995).  Using optical and {\it XMM-Newton} X-ray
observations, Mahdavi \etal\ (2005) identify this source as an
independent subgroup that has not yet interacted with, nor is bound
to, the main group.

\subsection{Unsharp-Masked Image} \label{sec:unsharp}

To better visualize faint surface brightness fluctuations,
particularly near the core, we made a (0.3--5.0~keV) unsharp-masked
image.  It was created by dividing the image smoothed with a
0.98\arcsec\ radius Gaussian by one smoothed with a 9.8\arcsec\
Gaussian.  The resulting image is shown in Figure~\ref{fig:unsharp1}.
Two bubbles are clearly seen to the north and southeast of the central
AGN, which are also detected in the radio (see
Figure~\ref{fig:unsharp2}).  
The surface brightness profiles in four sectors, two of which contain
the bubbles, are compared to the average profile in
Figure~\ref{fig:bubble_sb}.  This figure shows that the northern and
southern bubbles correspond to $\sim60$\% and $\sim40$\% deficits
respectively, as compared to both the average profile and the peaks at
larger radii which correspond to the bright X-ray rims.
In the radio, 6~cm observations clearly show a central core with two
radio lobes corresponding to the bubbles seen in the {\it Chandra}
images, with comparable radio flux from each lobe and the core (see
B2~1317+33 in Morganti \etal\ 1997).  In Figure~\ref{fig:unsharp2} we
overlay radio contours from VLA L-band 1.45 GHz data taken from the
VLA image archive over a close-up of the core in
Figure~\ref{fig:unsharp1}.  Radio emission fills the X-ray bubbles.

The unsharp-masked image also reveals
complex structure in the 
diffuse emission, most notably southwest of the AGN, with several
surface brightness depressions similar to those seen in the bubbles
(but without the surrounding bright rims),
all within the 
outer edge noted in Figure~\ref{fig:fullimg}.  There is no obvious
correlation between the radio observations and these 
other X-ray cavities.
A comparison of the net
counts in these depressions to those from adjacent regions shows that most
are statistically significant at the 2-3$\sigma$ level.
In the bubbles, the count rates in the rims are higher than those in the
central depressions by 4.6$\sigma$ and 2.3$\sigma$, for the northern
and southern bubbles respectively.  For one of the more significant
cavities to the southwest, the deficit is significant at 3.8$\sigma$.
Therefore, the statistical significance of the cavities is on the
order of that for the bubbles, which are seen in the radio and clearly
real features.
The large, dark region to the east, just outside the bubbles, is an
artifact of the relatively sharp drop-off in surface brightness in
this region, possibly indicating an edge in the central bright diffuse
emission (although the bright rims of the bubbles also contribute to
this deficit).

\section{Spectral Analysis} \label{sec:spec}

The X-ray image (Figure~\ref{fig:fullimg}) shows diffuse emission
associated with NGC~5098a, as well as fainter group emission filling
the field of view.  We generated a
temperature map as a guide for detailed spectral fitting to
disentangle the various components and study the structure seen in the
ICM.  We assume a galactic absorption of $N_H = 1.31\times 10^{20}$ cm$^{-2}$
throughout.

\subsection{Temperature Map} \label{sec:tmap}

The temperature map was derived using the same method as developed in
Randall \etal\ (2008; 2009). For each temperature map pixel, we
extracted a spectrum from a circular region containing 1000 net counts
(after subtracting the blank sky background).  The resulting spectrum
was fit in the 0.6 -- 5.0~keV range with an absorbed APEC model using
{\sc XSPEC}, with the abundance allowed to vary. The 
resulting temperature map, with X-ray surface brightness contours
overlaid, is shown in Figure~\ref{fig:tmap}.
Unfortunately, due to the small number of net counts, the extraction
regions for the temperature map pixels were relatively large.  Faint
regions had extraction radii 
on the order of 1.6\arcmin\ (71~kpc), while the brightest regions,
near the core of NGC~5098a, had radii of 7.9\arcsec\ (5.8~kpc). 
As a result, each pixel in the temperature map is highly
correlated with nearby pixels, and the temperature map is effectively
smoothed on large scales, particularly in regions far from the core.

Nevertheless, there are several interesting features in the
temperature map.  There is an elliptical clump of cool gas west of the
AGN, which appears to be doubly peaked to the north and south.  The
cool gas wraps around the AGN, but does not overlap it.  The outer
edge of this cool region roughly corresponds to the surface brightness
edge noted in Figure~\ref{fig:fullimg}, suggesting that this feature
is a cold front.  A 
long arm of cool gas extends east of the AGN and wraps around to the
north, with the outer edge connecting to the edge of the cool
elliptical region in the south.  The orientation and morphology of the
cool arm is very similar
to the arm seen in the surface brightness map
(Figure~\ref{fig:fullimg}), though it extends well beyond the apparent
outer boundary of the surface brightness arm.
We investigate this feature further in
\S~\ref{sec:dspec}.  
As in the surface brightness map, there is no
structure correlated with NGC~5098b.
As a test of the robustness of the temperature measurements in faint
regions, we re-fit the spectra for a few temperature map pixels near
the tip of the cool arm in the temperature map and varied the
background normalization by $\pm 10\%$.  We found no significant
effect on the resulting temperature.  The average radial temperature
structure is consistent with that found previously from analyses of
{\it Chandra} and {\it XMM-Newton} data (Xue \etal\ 2004; Gastaldello
\etal\ 2007), with $kT \approx 0.8$~keV near the core of
NGC~5098a, rising to $\sim1.3$~keV at 1.5-2\arcmin\ (67-89~kpc), and
dropping off at larger radii.

To search for correlation with the detailed structure shown in
Figure~\ref{fig:unsharp1}, we made a higher-resolution temperature map
of the central region of NGC~5098a, shown in
Figure~\ref{fig:tmap_core}, with the same X-ray surface brightness
contours as in Figure~\ref{fig:tmap}.  The smoothing scales are too
large to show any structure at the level of the X-ray cavities shown
in Figure~\ref{fig:unsharp1}, although the double-peaked nature of the
elliptical region, and its anti-correlation with the AGN, can be sen
more clearly.  It is also clear that the temperature in the inter-arm
region, between the arm's inner edge and the cooler central gas, is
consistent with the ambient temperature of $\sim 1.2$~keV.  This is
also consistent with what we find from detailed spectral fits (see
\S~\ref{sec:dspec}).

The structure of the temperature map, in particular the
cool spiral arm morphology and the offset of the cool gas from the central AGN,
supports our conclusion from \S~\ref{sec:img} that this is a disturbed
system, possibly due to an interaction with the nearby galaxy NGC~5098b.
In particular, these features are very similar to those seen from
sloshing of the central gas about the potential minimum (see Figure~7 in
Ascasibar \& Markevitch 2006, especially at 1.7 - 1.9~Gyr). We discuss
this possibility further in \S~\ref{sec:discuss}. 

\subsection{Detailed Spectra} \label{sec:dspec}

\subsubsection{Diffuse Emission} \label{sec:diffuse}

Based on the derived temperature map and the X-ray image, we
defined 7 regions for detailed spectral analysis.  A summary of the
spectral model for each region is given in
Table~\ref{tab:spectra}.  R1 is centered on the AGN and covers the
central projected 26\arcsec\ (19.2~kpc) of NGC~5098a.  R2 and R3
correspond to the southern and northern cool spots in the larger
elliptical region of cool gas seen in the temperature map,
respectively.  R4 covers the inner part of the cool arm seen in the
temperature map, while the adjacent region R5 is the hotter area
just outside of the temperature map arm (though R5 overlaps with the arm
seen in the surface brightness map).
The regions are shown overlaid on the X-ray image and temperature map
in Figure~\ref{fig:regs}, where in the left panel point sources have
been removed as in Figure~\ref{fig:smoothed}. 

A single-temperature thermal model was an adequate fit to each region
considered in Table~\ref{tab:spectra}.  The best-fitting temperature
for the ``inter-arm'' region seen in the temperature map (R5) was
greater than that in the adjacent region R4, with a significance of
3.7$\sigma$.  The best-fitting abundance was also higher in R5, by more
than a factor of two.  As a test, we re-fit the spectrum from R5 with
the abundance fixed to the best-fitting value from region R4 (row~6 in
Table~\ref{tab:spectra}).  The chi-squared per degree-of-freedom was
significantly worsened by fixing the abundance, and the best-fitting
temperature was still found to be higher, with a significance of
2.3$\sigma$.  This suggests that the arm structure seen in the
temperature map in the vicinity of R4 and R5 is real, and not an
artifact of smoothing.

As noted in \S~\ref{sec:tmap}, the cool arm seen in the temperature
map extends beyond the outer edge of the arm in the surface
brightness map.  R6 and R7 were defined to evaluate the
significance of the temperature difference in these
regions. Table~\ref{tab:spectra} shows that the best-fitting
temperature in R7 is higher than that in R6 with high significance
(greater than 6$\sigma$).  This confirms that the emission near the
edge of the FOV is hotter northwest of the AGN as compared to the
northeast, as suggested by the arm feature in the temperature map,
independent of smoothing effects.

\subsubsection{The Surface Brightness Edges} \label{sec:edges}

The surface brightness edges indicated in Figure~\ref{fig:fullimg} are
similar to features arising from contact discontinuities, or ``cold
fronts'', seen in simulations and observations of galaxy clusters and
groups (\eg, Ascasibar \& Markevtich 2006; Dupke \etal\ 2007;
Gastaldello \etal\ 2009).  We therefore extracted spectra across these
edges to search for temperature jumps, which are characteristic of
cold fronts.  Spectra were extracted from the same semi-annular
regions shown in Figure~\ref{fig:sect}, but with fewer (larger) radial
bins, such that each bin contained $\sim500$ net counts.  The
resulting spectra
were fit with a single temperature APEC model, with the abundance
allowed to vary.  The resulting temperature profiles are shown in
Figure~\ref{fig:ktprof}.  Each profile shows a clear jump at the
location of the apparent surface brightness edge, identifying these
features as cold fronts.  These features are discussed in more detail
in \S~\ref{sec:fronts}.  We note that, for
the northeastern profile, the elevated temperature of the innermost bin (which
overlaps with region R5 in Figure~\ref{fig:regs}) and the lower
temperatures at large radii (as compared to the southwestern profile)
are consistent with what is shown the temperature map (Figure~\ref{fig:tmap}).

\subsubsection{The Central Source} \label{sec:agn}

We extracted a spectrum for the central source in NGC~5098a using an
aperture with a radius of 1.8\arcsec.  The background was determined
locally from an annulus centered on the source with inner and outer
radii of 1.8\arcsec\ and 4.0\arcsec, respectively.  This gave 264 net counts
in the 0.6 -- 7.0~keV band.  The spectrum was fit with an absorbed
power-law, giving a best-fitting photon index of $1.93 \pm 0.16$, which
is typical for a radio galaxy (\eg, Sambruna \etal\ 1999).  The source
has an unabsorbed flux of $4.2 \times
10^{-14}$~ergs~cm$^{-2}$~s$^{-1}$ and X-ray luminosity of $1.4 \times
10^{41}$~ergs~s$^{-1}$ in the 0.6 -- 7.0~keV energy band.  We also fit
the spectrum with the {\sc XSPEC} intrinsic absorption model ZWABS and find
that the spectrum is consistent with no internal absorption.

\section{Discussion} \label{sec:discuss}

\subsection{Structure of the Cold Fronts} \label{sec:fronts}

The {\it Chandra} image (Figure~\ref{fig:fullimg}) shows a
brightness edge almost completely encircling NGC~5098a.  It is
sharpest and closest in the west/southwest, continues on to the south,
and spirals out to form the outer boundary of the surface brightness
arm in the northeast.  Spectral analysis revealed temperature jumps
across each edge,  indicating that these edges are cold fronts
(\S~\ref{sec:edges}). As such, we expect
there to be a density discontinuity located at each edge.  To measure
the amplitude of these density jumps, we extracted the {\it Chandra}
0.6--5.0~keV surface brightness profile in two sectors, one to the
southwest and another to the northeast. The regions were defined such that the
radii of curvature matched those of the edges.  In each case, the
center of curvature was reasonably close to the peak of the diffuse
X-ray emission, near the central AGN. The extraction regions are shown in
Figure~\ref{fig:sect}.

The resulting emission measure profiles are shown in
Figure~\ref{fig:profs} (distance is measured from the center of
curvature of the apparent edge). These profiles were fit with a
spherical gas density model consisting of two power laws.  The free
parameters were the normalization, the inner ($\alpha$) and outer
($\beta$) slopes, the position of the density discontinuity ($r_{\rm
  break}$), and the amplitude of the jump ($A$).  The temperature and
abundance for each bin were determined from the fits to the projected
emission from the coarser bins shown in Figure~\ref{fig:ktprof}, as there were
insufficient counts to perform deprojected spectral fits.  The profile
from the southwest sector has the typical shape expected for a contact
discontinuity. Fitting our model, we find $\alpha =
-1.15^{+0.07}_{-0.06}$,  $\beta = -0.88^{+0.06}_{-0.06}$, $r_{\rm
  break} = 31^{+2}_{-1}$~kpc, and $A = 2.13^{+0.21}_{-0.17}$.  In
the northeast, the density jump is less pronounced, but still
significant within the errors:  $\alpha = -0.70^{+0.11}_{-0.12}$,
$\beta = -1.47^{+0.16}_{-0.15}$,  $r_{\rm break} = 56^{+3}_{-5}$~kpc ,
and  $A = 1.45^{+0.22}_{-0.17}$. 
We note that the presence of the long trail of cool gas seen in the
northeast in Figure~\ref{fig:tmap} could contaminate the temperature
profile shown in Figure~\ref{fig:ktprof} ({\it right}), and therefore affect our
fits to the density jump in this region.  However, the emissivity is
only weakly dependent on the temperature (which does not strongly
vary in this region), so the integrated emissivity
profile we fit to is dominated by surface brightness variations.
Furthermore, if we were able to remove any contribution from the cool
arm beyond the surface brightness edge we would raise the overall
temperature in this region, which would increase the size of the
temperature jump in  Figure~\ref{fig:ktprof} ({\it right}) and only
strengthen our conclusion that this surface 
brightness edge corresponds to a cold front.
For both the northeastern and southwestern edges, the two power-law model was a 
much better fit to the data than a single power-law or beta model.
The best fit models are plotted in Figure~\ref{fig:profs}.

\subsection{The Dynamical State of NGC~5098} \label{sec:state}

Based on temperature profiles and fits to the surface brightness
profiles (see \S~\ref{sec:fronts} and \S~\ref{sec:edges}), the edges
shown in Figure~\ref{fig:fullimg} are identified as cold fronts.  Such
contact discontinuities are expected to arise during a merger,
either at the leading edge of a remnant gas core, or from
the sloshing of gas around the potential minimum after it is perturbed
by an off-axis encounter with a secondary object (see Markevitch \&
Vikhlinin 2007 for a review).  There are several reasons to identify these
features as sloshing fronts.  The centers of curvature of the features
are roughly centered on the X-ray brightness peak, as expected for
sloshing fronts.  The density jumps are modest, in the 1.1--2.1 range,
in contrast to the larger jumps seen in remnant core cold fronts (\eg,
Markevitch \etal\ 2002; Randall \etal\ 2009).  The spiral structure,
seen in both the X-ray surface brightness and temperature maps
(Figures~\ref{fig:fullimg}~\&~\ref{fig:tmap}), is characteristic of
sloshing fronts in existing observations and simulations (Ascasibar \&
Markevitch 2006; Dupke \etal\ 2007; Gastaldello \etal\ 2009). The
apparently nearby galaxy NGC~5098b provides a likely interaction
candidate to initiate sloshing in the core of NGC~5098a.  Finally, and
most compellingly, the temperature map (Figure~\ref{fig:tmap_core})
clearly shows an offset between the cool central gas and the central
AGN (and the centroid of the optical emission), which presumably lies
at the local minimum of the potential.  The structure is very similar
to that observed in simulations of gas sloshing (\eg, Figure~7 in
Ascasibar \& Markevitch 2006)

As noted in \S~\ref{sec:tmap}, one potential problem with this picture
is that the cool arm seen in the temperature map
(Figure~\ref{fig:tmap}) extends well beyond the apparent outer boundary
of the surface brightness arm.  Detailed spectral fits, described in
\S~\ref{sec:dspec}, show that the arm is likely a real feature, and
not an artifact of the smoothing inherent in the temperature map.
While it is possible that the arm represents sloshed gas from the core
of NGC~5098a, the scales involved make this explanation seem unlikely.
To explain this feature, we note that NGC~5098b does not show any
associated emission in X-rays, implying that this system has been
stripped of its gas.  The arm seen in the temperature map may be gas
stripped from NGC~5098b as it fell into the group from the
northeast and interacted with the ambient ICM.  The orientation of the
arm is consistent with the sloshing interpretation, as the direction of
angular momentum implied by the winding direction of the surface
brightness spiral arm morphology is consistent with a perturber that
has approached from 
the northeast and passed east of NGC~5098a (in projection).  In this
scenario, the apparent connection between the outer temperature map arm and
the inner surface brightness arm is a projection effect, which may
explain the less than perfect correspondence between these features
near the tip of the surface brightness arm.  
Furthermore, it is possible that NGC~5098b, after being ram pressure
stripped and passing east of NGC~5098a, circled around NGC~5098a (not
necessarily in the plane of the sky) and
is now moving roughly to the east, creating a subtle conical wake in the
diffuse emission, similar to those seen in the first and last panels
of Figure~21 in Ascasibar \& Markevitch (2006).  The surface
brightness arm could be a manifestation of the wake, rather than
cool gas sloshed from the core of NGC~5098a, which would explain the
lack of correlation with the temperature map.
Unfortunately, we do not
have enough net counts to test in detail either the interpretation of the
extended cool arm in the temperature map as stripped gas from
NGC~5098b, or the possibility that the surface brightness arm is a
conical wake generated by the return passage of NGC~5098b.  We note
only that the current data are consistent with either interpretation.

As shown in Figure~\ref{fig:unsharp2}, the central bubbles are swept
back to the east of the central AGN in projection.  
This could be the result of ram pressure due to 
relative motion between the host galaxy and the ICM either due to the 
galaxy's motion within the group or bulk motion of the gas.
Alternatively, the jets may have encountered and been
deflected by inhomogeneities
in the ICM.
We note that, in simulations, sloshing fronts are not static
features. Rather, gas flows along the spiral arm, from the outer regions into
the center (Figure~7 in Ascasibar \& Markevitch 2006).  In this system, we
therefore expect gas to be flowing inwards from the east, around the
the south, and approaching the central AGN from the west.  It is
therefore possible that the bulk motions in the gas which sweep
the central bubbles to the east are due to the velocity field set up
in the sloshing front, though we cannot 
distinguish this possibility from the others previously mentioned.

\subsection{The X-ray Cavities} \label{sec:cavities}

\begin{figure}
\plotone{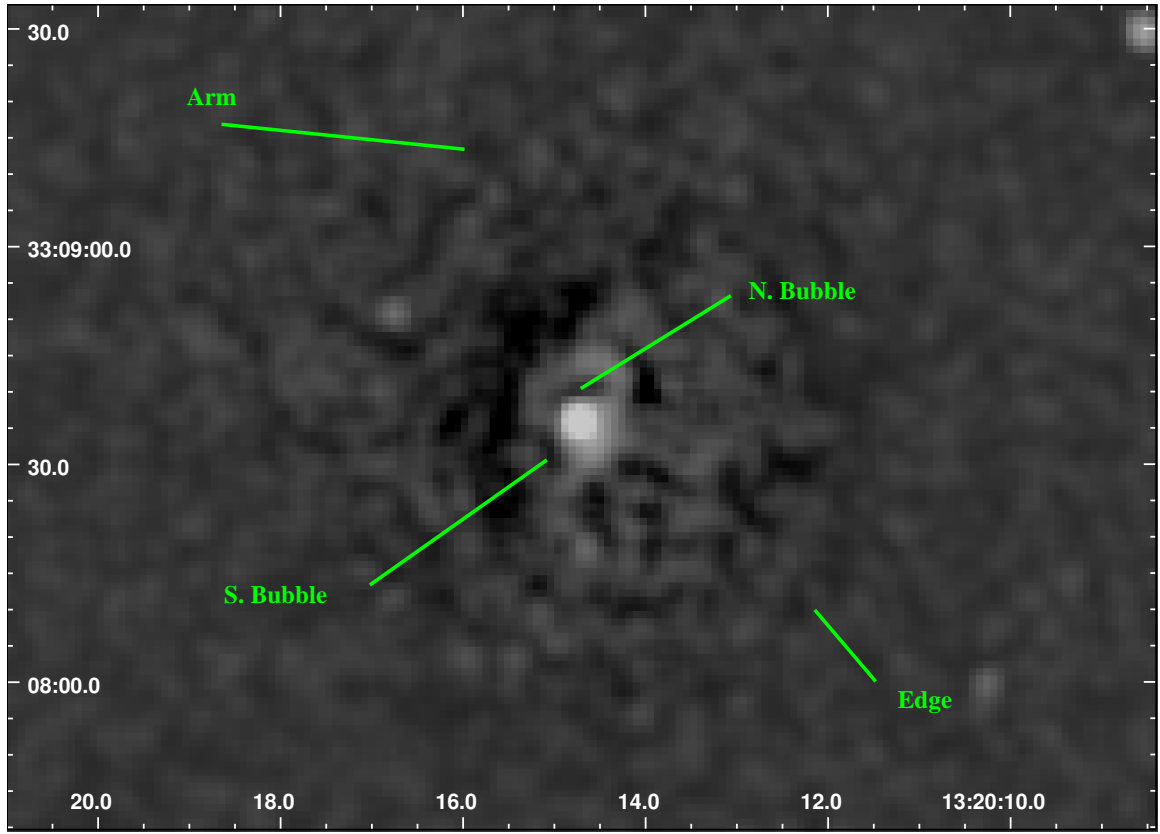}
\caption{
  An unsharp-masked 0.3--5~keV image of the central region of NGC~5098a.  X-ray
  cavities, or bubbles, encompassed by bright rings of emission are
  clearly seen to the north and southwest of the AGN.  
  The bubbles are filled with radio-emitting plasma (see
  Figure~\ref{fig:unsharp2}).
  The
  unsharp-masking reveals other fine structure and surface brightness
  depressions, particularly in the southwest (all within the surface
  brightness edge shown in Figure~\ref{fig:fullimg}).
  \label{fig:unsharp1}
}
\end{figure}

\begin{figure}
\plotone{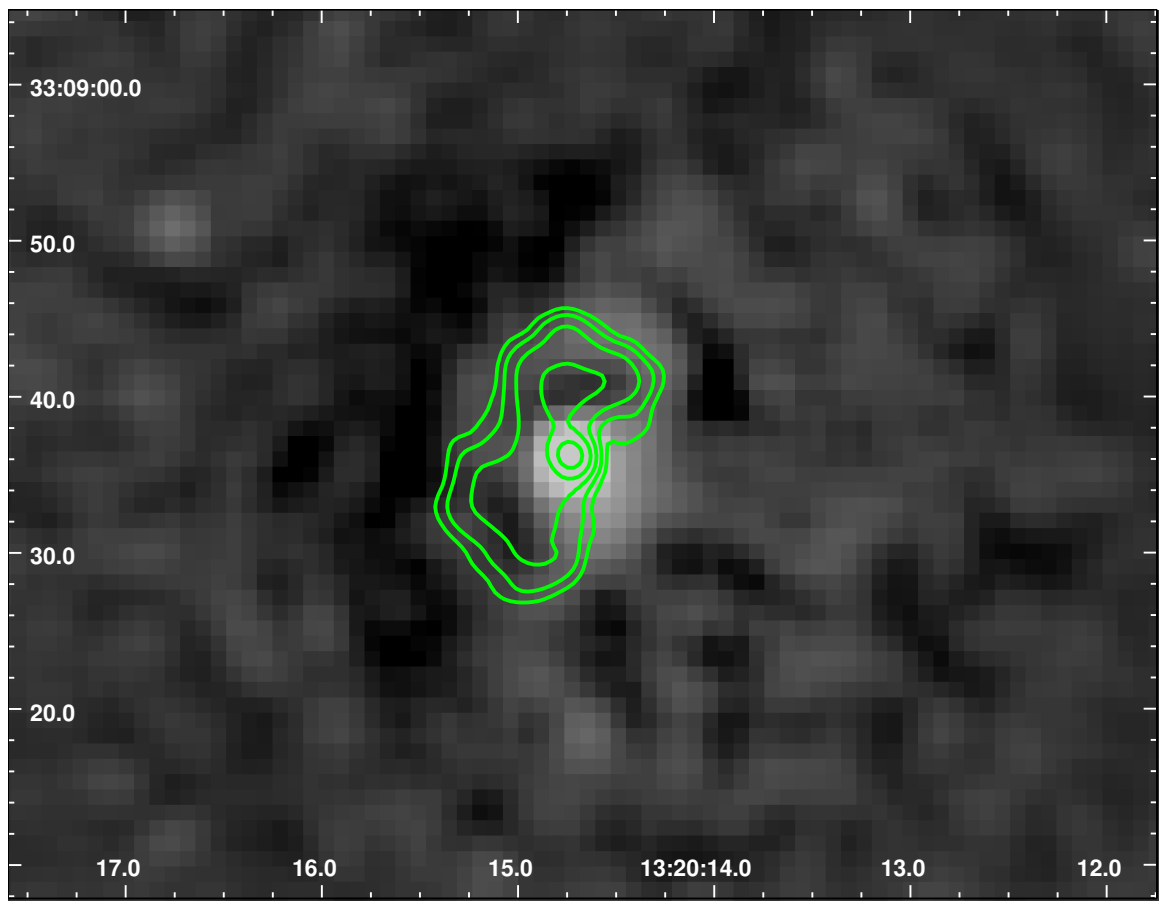}
\caption{
  A close-up of the bubbles shown in Figure~\ref{fig:unsharp1}.  The
  logarithmically-spaced contours were generated from 1.45~GHz VLA
  L-band images taken from the VLA data archive.  Radio emission fills
  the X-ray bubbles.
  \label{fig:unsharp2}
}
\end{figure}

\begin{figure}
\plotone{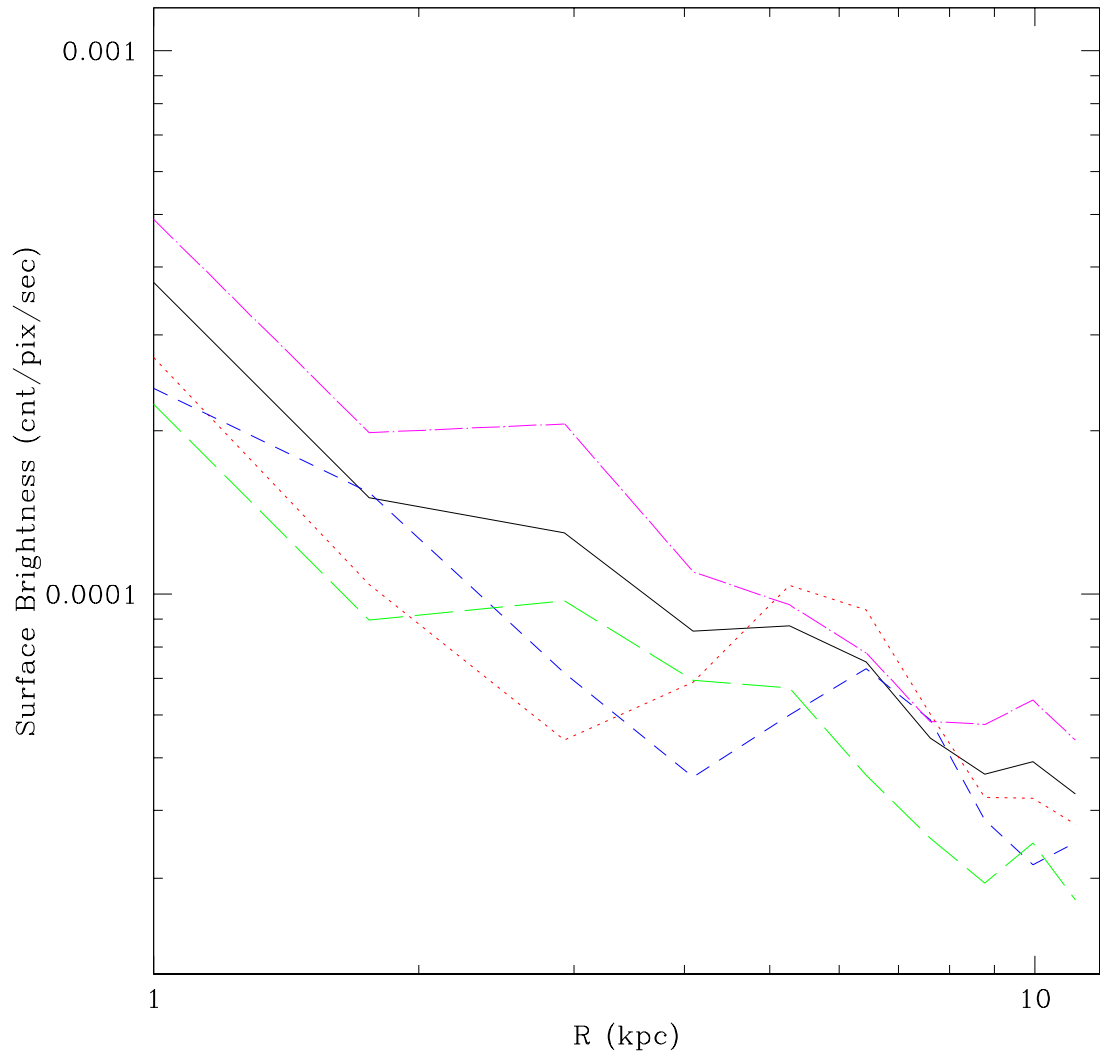}
\caption{
The 0.6-5.0~keV surface brightness profile in four sectors containing
the northern bubble (red dotted), the southern bubble (blue
short-dashed), the region to the east (green long-dashed), and the
region to the west (magenta dot-dashed) as compared to the azimuthal
average (black solid).  Error bars have been omitted for clarity.
The peaks at $\sim6$~kpc correspond to the bright bubble rims, and
the deficits at 3-4~kpc to the bubbles themselves.  The
overall east-west asymmetry is also evident.  Pixels are
0.98\arcsec$\times$0.98\arcsec.
\label{fig:bubble_sb}
}
\end{figure}

\begin{figure}
\plotone{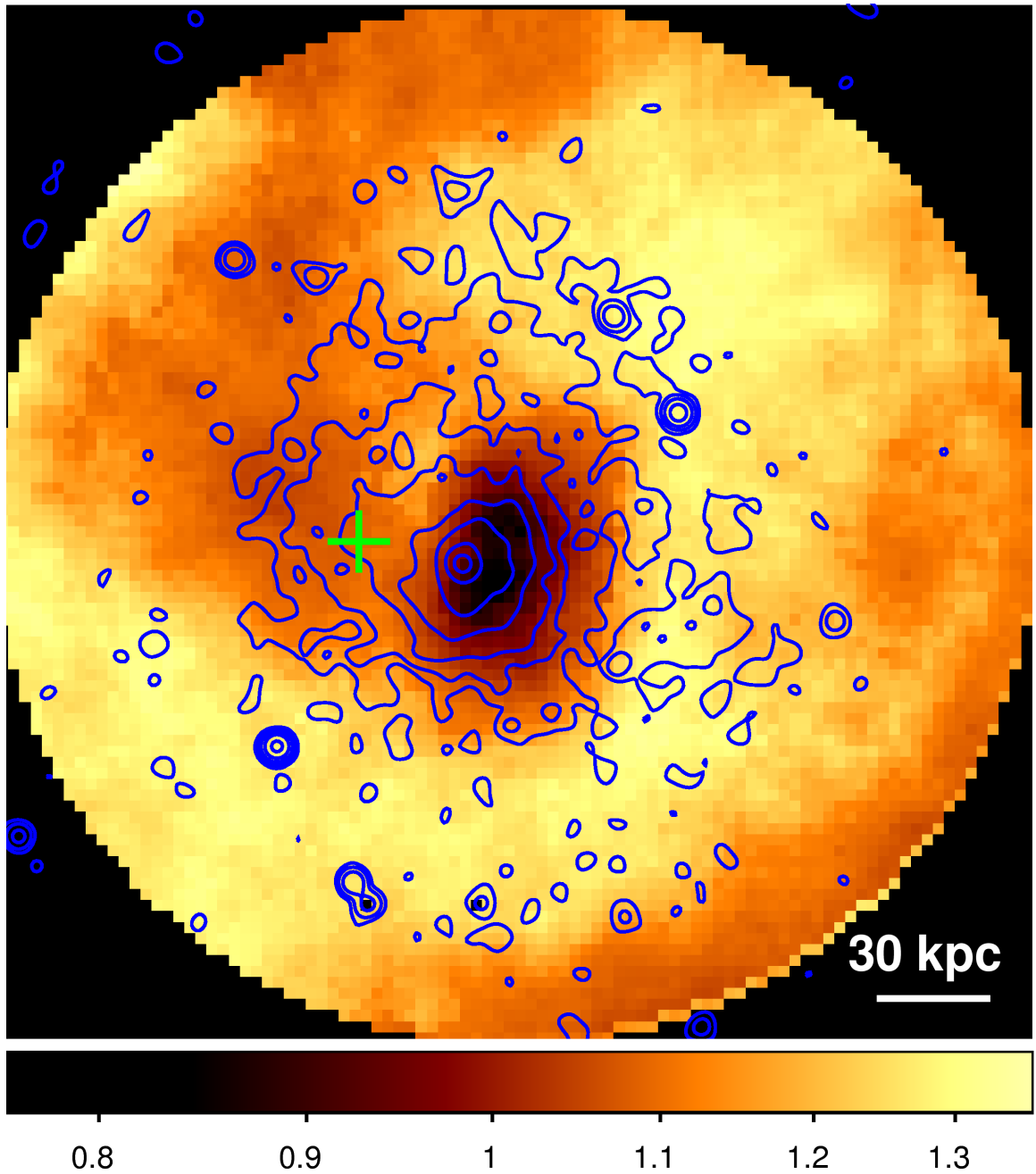}
\caption{Temperature map derived from the ACIS-S3 data, with
  {\it Chandra} X-ray logarithmic surface brightness contours
  overlaid. The green cross indicates the optical position of
  NGC~5098b.  The
  color-bar gives the temperature in keV.
\label{fig:tmap}
}
\end{figure}

\begin{figure}
\plotone{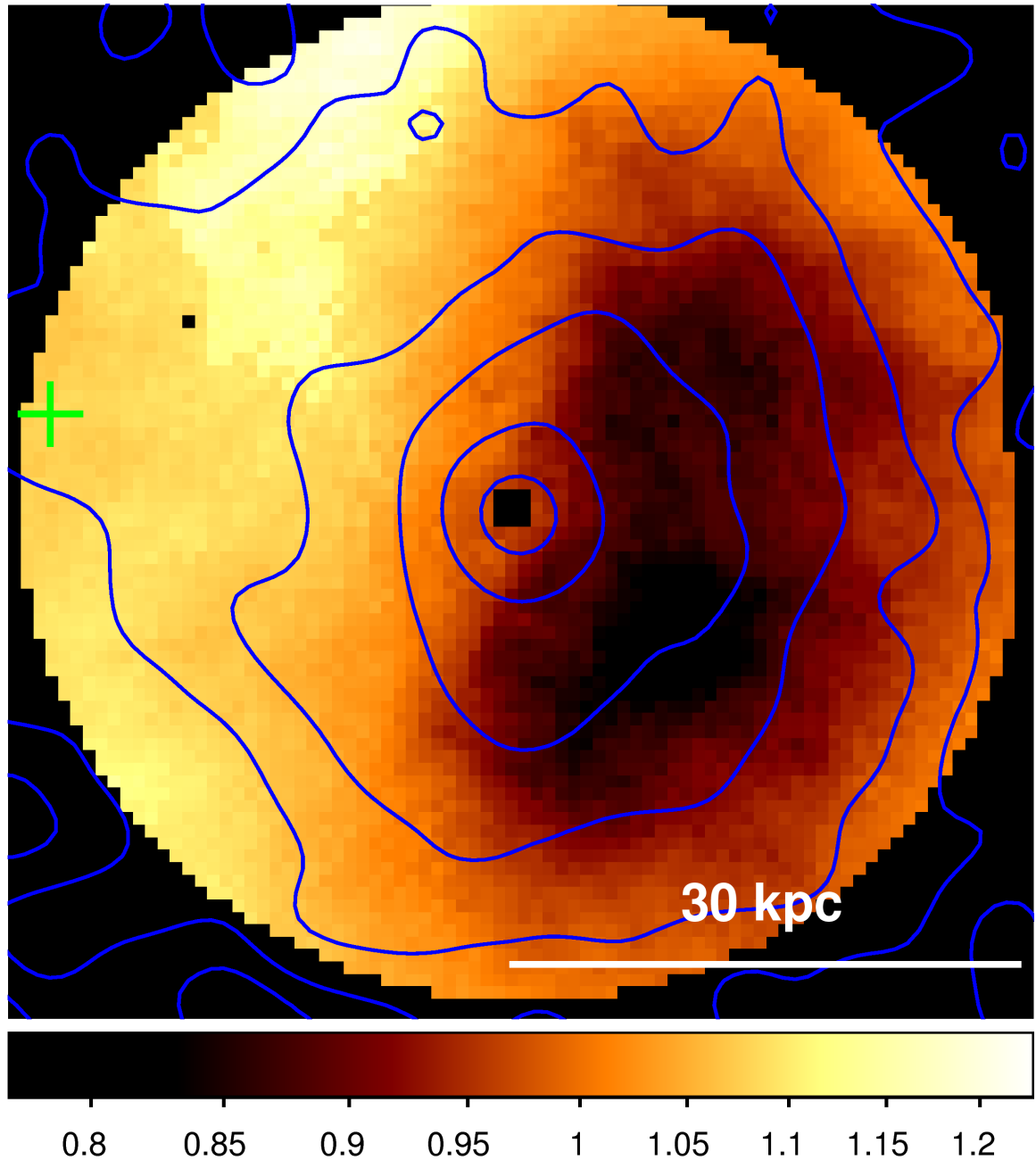}
\caption{A higher resolution temperature map of the central region of
  NGC~5098a, with the same X-ray surface brightness contours overlaid
  as in Figure~\ref{fig:tmap}.   The green cross indicates the optical
  position of NGC~5098b. The elliptical cool region to the west
  is split into two cool spots, one to the north and one to the
  south. The holes in the temperature map indicate pixels
  that were completely contained within an excluded source region
  (\eg, at the central AGN).
\label{fig:tmap_core}
}
\end{figure}

\begin{figure}
\plotone{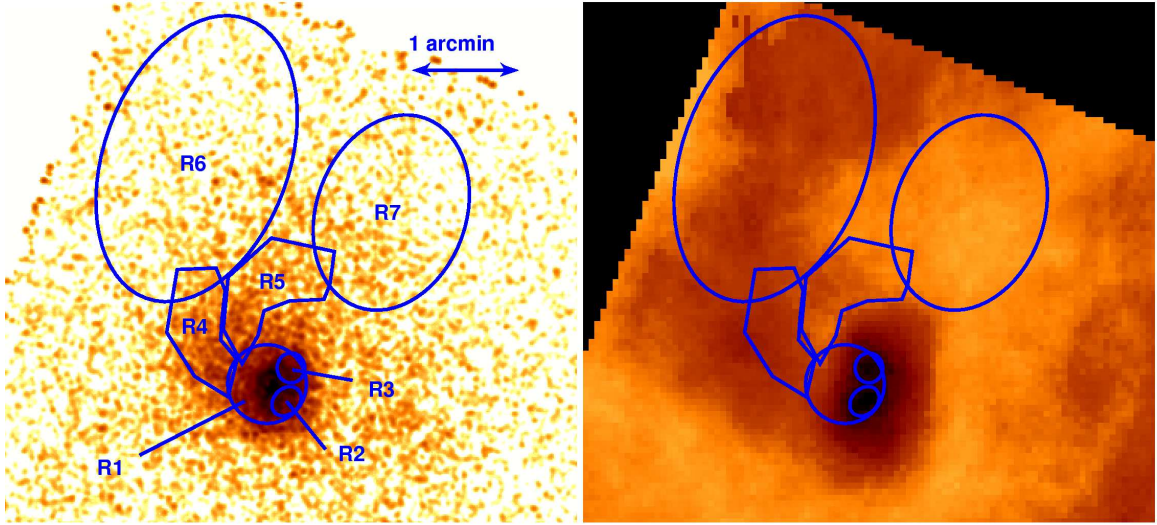}
\caption{
The regions fitted in Table~\ref{tab:spectra} overlaid on the 0.6 --
5.0~keV {\it Chandra} image ({\it left}) and temperature map ({\it
  right}).  The image has been
smoothed with a 6\arcsec\ radius gaussian, and the point sources have
been removed by filling in source regions as described in the text
(\S~\ref{sec:dspec}).
\label{fig:regs}
}
\end{figure}

\begin{figure}
\plottwo{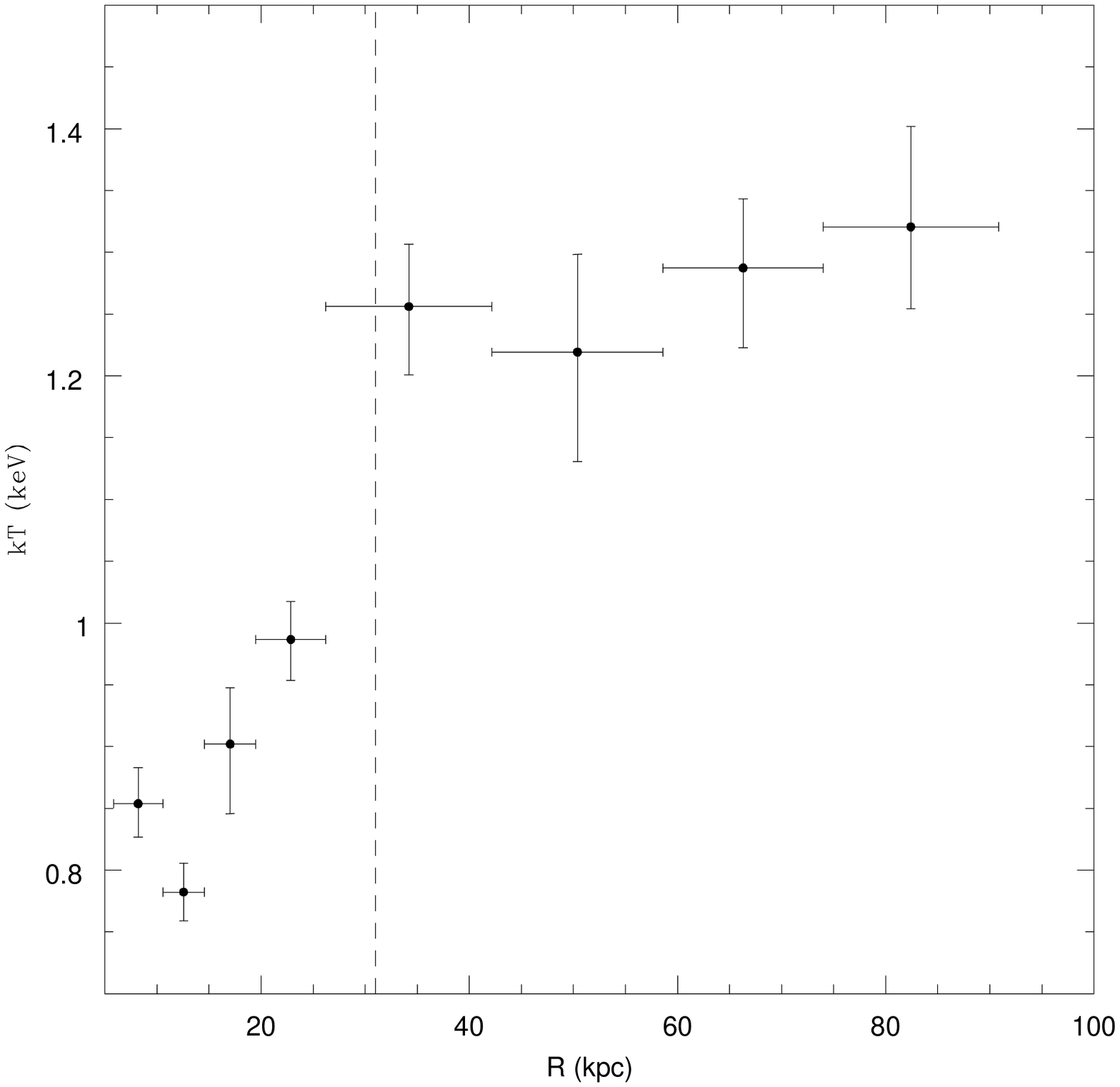}{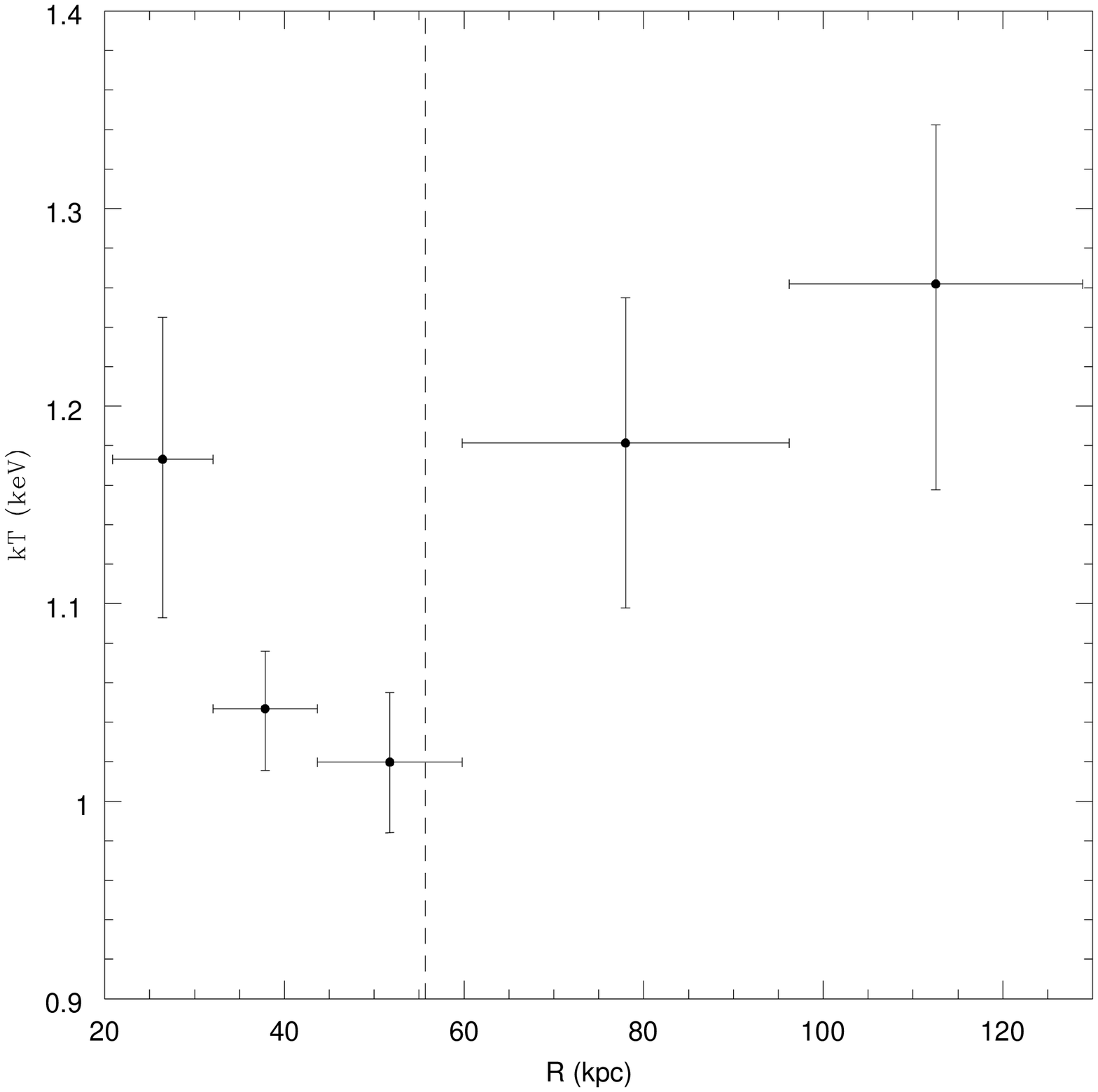}
\caption{
  Temperature profiles to the southwest ({\it left}) and northeast
  ({\it right}) for the semi-annular areas shown in
  Figure~\ref{fig:sect} (but with fewer, larger bins).  The dashed
  lines mark the positions of the density jumps calculated in
  \S~\ref{sec:fronts}.
  \label{fig:ktprof}
}
\end{figure}

\begin{figure}
\plotone{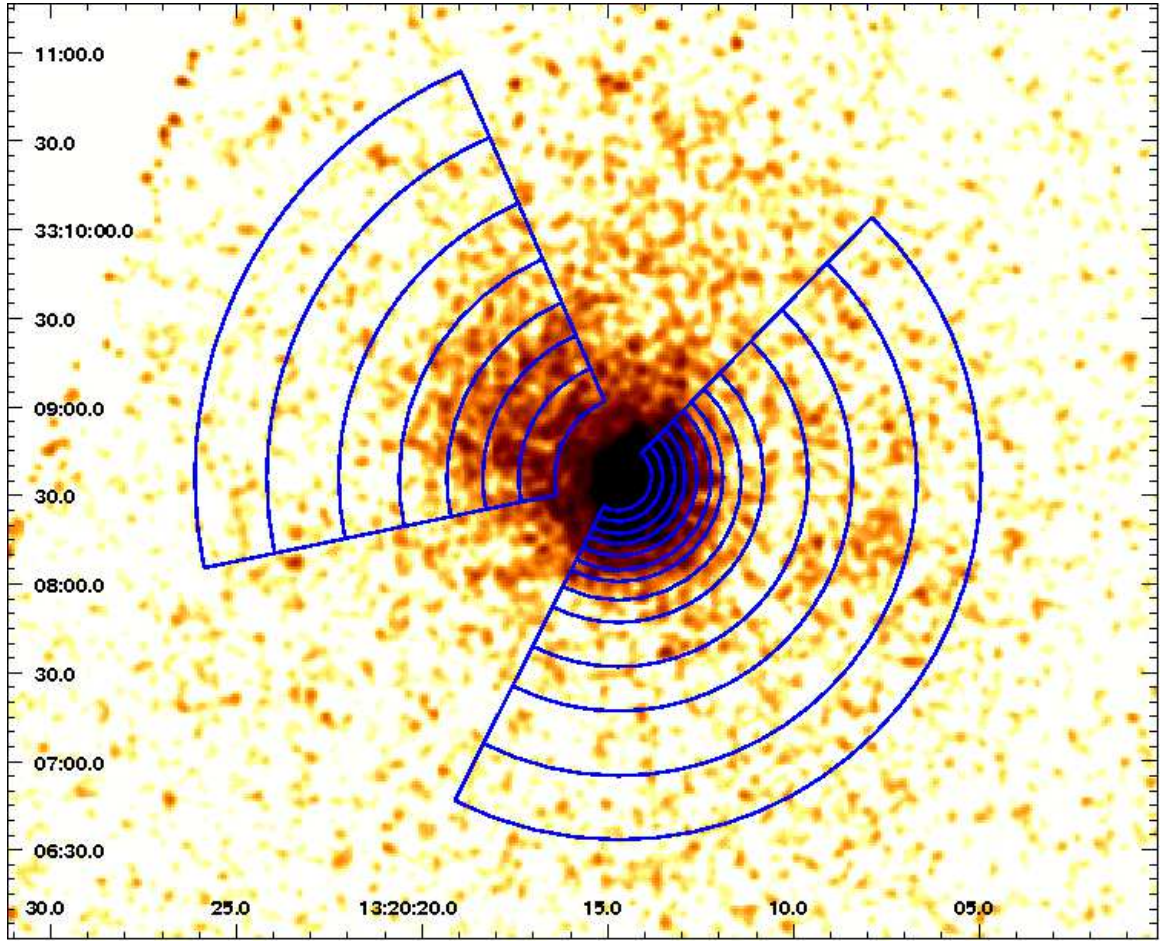}
\caption{
Regions used to generate the projected emission
measure profiles shown in Figure~\ref{fig:profs} overlaid on the
background subtracted, exposure corrected, smoothed, 0.6 -- 5.0~keV
{\it Chandra} image (with point sources removed).
\label{fig:sect}}
\end{figure}

\begin{figure}
\plotrtwo{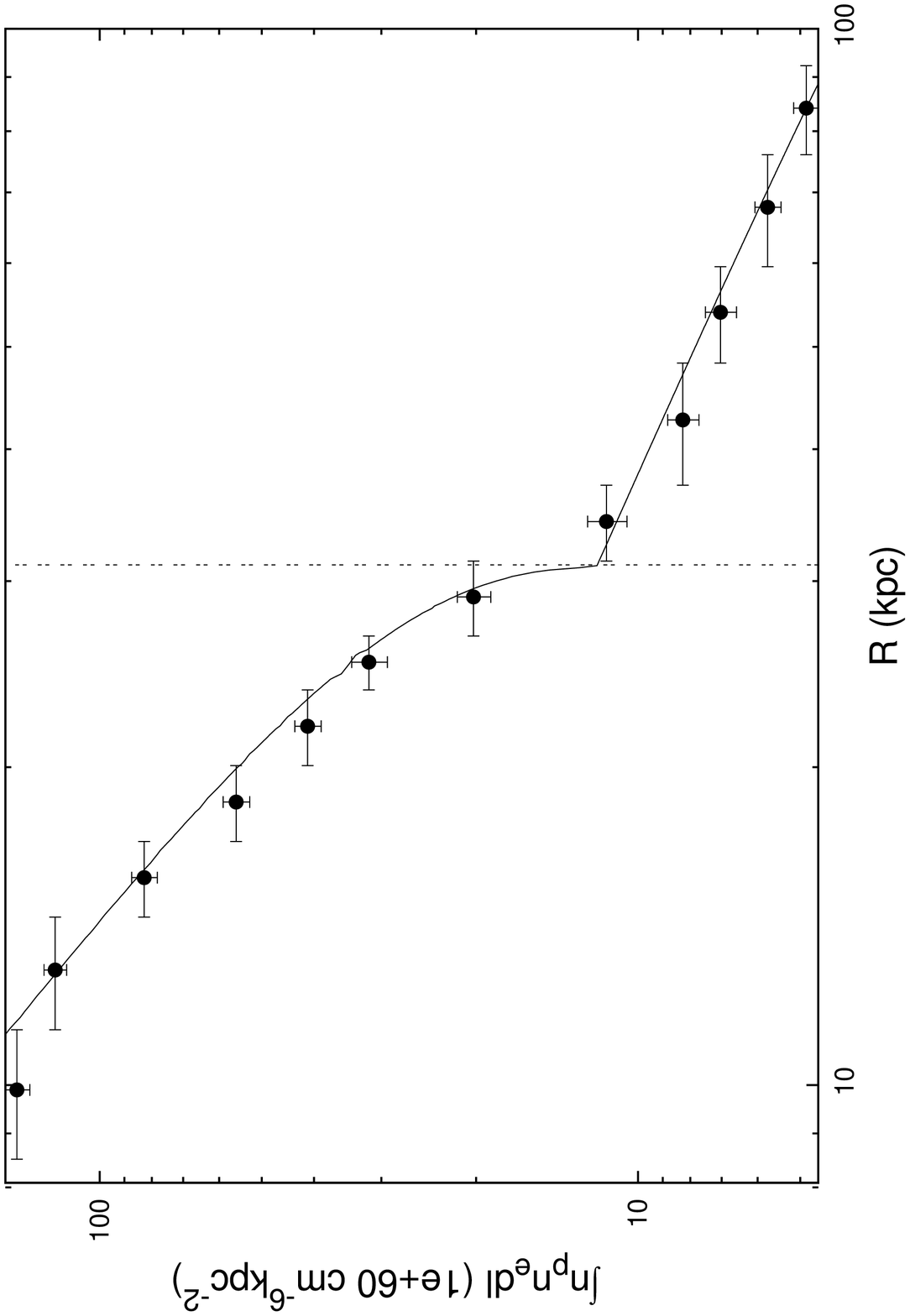}{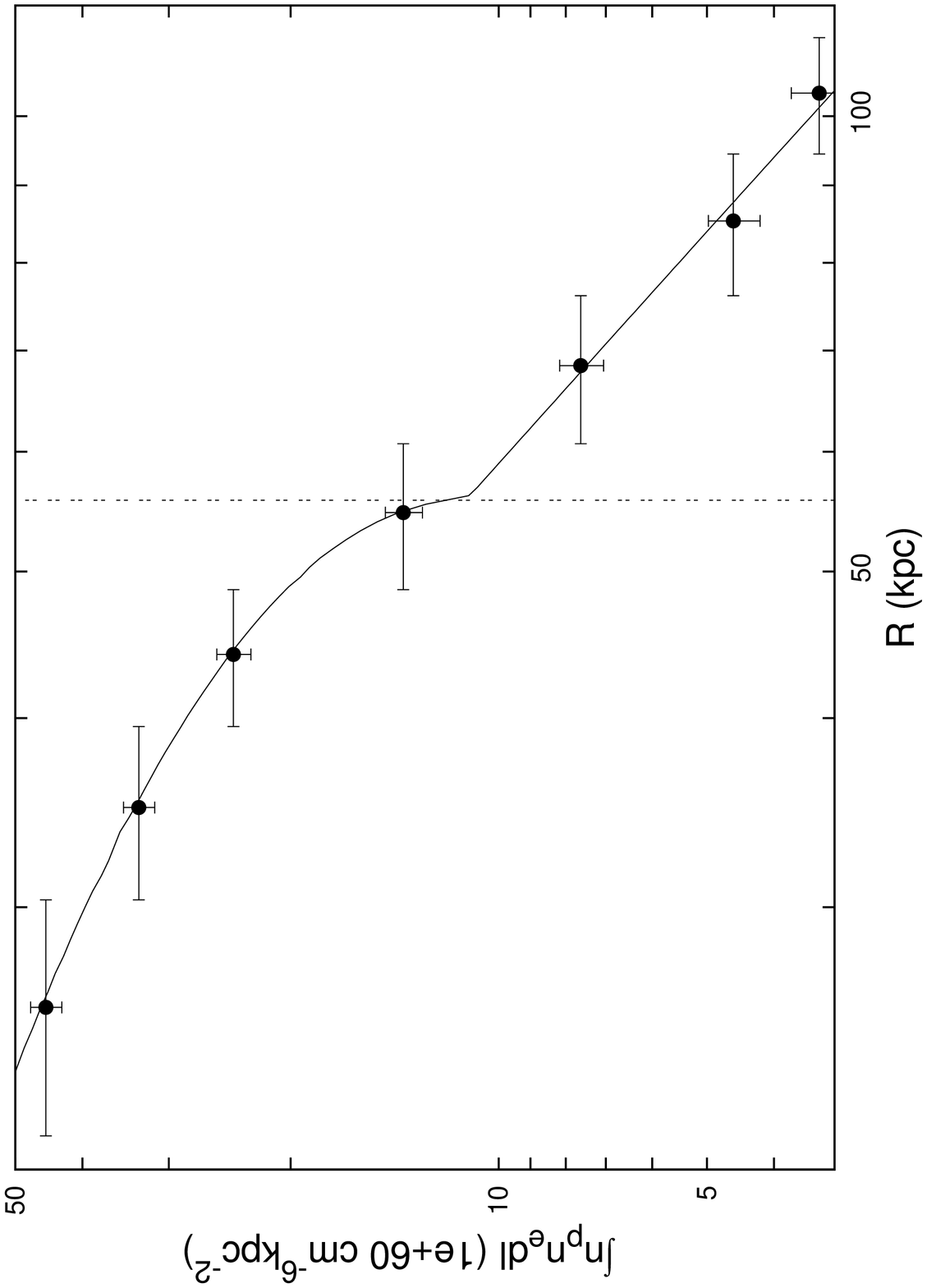}
\caption{
{\it Left Panel:} Integrated emission measure profile extracted from
the southwestern region 
shown in Figure~\ref{fig:sect}.   The x-axis gives
the radius from the apparent center of curvature defined by the
feature.  The best fit two power law density jump model is given by the
solid line. The vertical dashed line shows the best fit location of
the density discontinuity.  {\it Right Panel:} Same for northeastern
region in Figure~\ref{fig:sect}.
\label{fig:profs}}
\end{figure}

The X-ray surface brightness (Figure~\ref{fig:fullimg}) and
unsharp-masked (Figure~\ref{fig:unsharp1}) images show several
statistically significant
cavities in the central region of NGC~5098a.  In particular, there are
two distinct bubbles (seen as X-ray deficits surrounded by rims of
bright emission) just north and southeast of the central AGN.
The bubbles are the only cavities which correlate with existing radio
observations (see 
\S~\ref{sec:unsharp} and Figure~\ref{fig:unsharp2}), and we focus on these
features first.

The bubbles are morphologically similar to features seen in X-ray
observations of other galaxies, groups, and clusters.
These structures are formed when AGN jets push into the local ICM,
evacuating cavities, and often creating bright rims of X-ray emission
from the displaced gas.  
In more evolved remnant cavities the rims tend to be cooler and more
dense than the
nearby ambient ICM, as in
Abell~2052 (Blanton \etal\ 2003) and Perseus (Fabian \etal\ 2006),
whereas in more recent outbursts they often show higher temperatures
associated with shocks,
as in NGC~4552 (Machacek \etal\ 2006), Hercules~A (Nulsen \etal\
2005), and Centaurus~A (Croston \etal\ 2009).
To test for a temperature difference in the rims, we
extracted spectra from the northern (brighter) bubble rim and a similar region
just outside the rim, subtracted off spectra from local background
regions, and fit each with an absorbed APEC model with the 
abundance fixed at 30\% solar.  We find best-fitting temperatures of
$kT_{\rm rim} = 0.978^{+0.085}_{-0.096}$ for the rim and $kT_{\rm outside-rim}
= 1.085^{+0.085}_{-0.198}$ just outside the rim.  Although the
best-fitting temperatures indicate that the rim is somewhat cooler,
the difference is not statistically significant.

The proximity of the bubbles to the central AGN, and their relatively
small physical size, suggest that they are
currently forming as the cavities are inflated by the AGN.  If this is
the case, one might expect the total pressure in the cavities to
be on the order of that in the rims and the surrounding ICM, possibly
larger if the cavities are driving shocks.
We can estimate the
pressure in the X-ray emitting gas in the northern rim using the fit to
the spectrum to estimate the temperature and density of the gas.
Assuming an edge-on oblate spheroidal geometry for the bubble (with
semi-major and minor axes of 3~kpc and 1.6~kpc, respectively), we find an
electron density of $n_e \approx 0.03$~cm$^{-3}$ and a pressure of
$P_{\rm rim} \approx 4.5 \times 10^{-11}$~dyne~cm$^{-2}$.  To estimate
the pressure inside the bubble, ideally one would like flux
measurements from radio observations at multiple frequencies to
determine the spectral index of the relativistic particle population.
Under the assumption of equipartition, the radio pressure at minimum
energy is given by
\begin{equation}\label{eq:prad}
P_{\rm rad} = \frac{B^{2}_{\rm min}}{8 \pi} + \frac{4 \, E_{\rm min}}{7 \,
  \phi \, V}, 
\end{equation}
where the magnetic field at minimum total energy
\begin{equation}\label{eq:bmin}
B_{\rm min} = \left[ 6 \pi (1 + k)\, c_{12}(\alpha, \nu_1, \nu_2)\, L\, \phi^{-1}\, V^{-1}
\right]^{2/7}, 
\end{equation}
and the minimum total energy
\begin{equation}\label{eq:emin}
E_{\rm min} = \frac{7}{4} \left[ \frac{1}{6 \pi} \phi\, V (1 + k)^{4/3}
  (c_{12}(\alpha, \nu_1, \nu_2) \, L)^{4/3} \right]^{3/7}.
\end{equation}
In these relations, $k$ is the ratio of proton to electron energies,
$V$ is the volume of the emitting region, 
$\phi \approx 1$ is the volume filling factor, $L$ is
the radio luminosity at a given frequency, and $c_{12}(\alpha, \nu_1,
\nu_2)$ is a parameter (tabulated in Pacholczyk 1970) that depends on
the spectral index $\alpha$ 
and the lower and upper cut-off frequencies, which we take to be
$\nu_1 = 10$~MHz and $\nu_2 = 10$~GHz. We assume $k \approx 1$, which
is expected for a young source since it has not yet had time to
entrain material from the ICM.
Useful expressions from these
relations are given by Govoni \& Feretti (2004).
Morganti \etal\ (1997) give 6~cm fluxes of  8.7, 10.3, and 7.1 mJy for the
core, the northern lobe, and the southeastern lobe, respectively.
Unfortunately, from the literature we were only able to find radio flux
measurements for the resolved AGN and lobes at a single frequency, so
the value of $\alpha$ inside the lobes is unknown.  As a rough
estimate, we assume that $\alpha = -1.6$, consistent with results from
similar X-ray bubbles in Abell~2052 (Zhao \etal\ 1993; see B\^{i}rzan
\etal\ 2008 for indicies for several sources, though they are given
for the total source, not just the lobes, and hence are expected to be
steeper).  We find
an equipartition magnetic field strength of $B_{\rm eq} \sim40~\mu$G and a
minimum radio pressure 
of $P_{\rm rad} = 1.5 \times 10^{-10}$~dyne~cm$^{-2}$, more than three times the X-ray
pressure in the rims.  However, this result is sensitive to the assumed
value of $\alpha$: if we instead take $\alpha = -1$, which is closer
to the value for the bubbles in the Perseus cluster (Pedlar \etal\
1990), we find  $P_{\rm 
  rad} = 2.6 \times 10^{-11}$~dyne~cm$^{-2}$, {\it less} than the
X-ray pressure in the rims.  
Morganti \etal\ (1997) find that the ratio of flux from the northern
radio lobe to the total flux from both lobes and the central core is
$f \approx 0.43$ at 6~cm (5~GHz).
The NRAO/VLA Sky Survey (NVSS, Condon \etal\ 1998) gives a total flux
of 82.9~mJy at 1.4~GHz. 
Assuming that $f \approx 0.43$ at 1.4~GHz, we find 
$\alpha \approx -1$.  We note that this is an upper limit on $\alpha$, 
since the core is expected to be brighter at higher frequencies (\eg,
the lobes are 
not seen in 8.49~GHz X-band images from the VLA archive, whereas the
AGN is clearly visible), so that $f$
is a decreasing function of frequency.  We therefore conclude that $P_{\rm 
  rad} > 2.6 \times 10^{-11}$~dyne~cm$^{-2}$ is a hard lower limit,
and the radio pressure in the northern bubble is on the order of the
pressure in the surrounding X-ray emitting rims.
The fact that pressures are roughly equal under the assumption that $k
= 1$, and that the bubbles are physically close to the AGN, suggest
that the bubbles are young features, possibly currently being inflated by
the central AGN.
Follow-up high-resolution radio observations, along with deep X-ray
observations, would be of interest for a better comparison of the
relative pressures in the rims and bubbles, and of the X-ray
temperature in the rims and the ambient ICM.

The unsharp-masked image (Figure~\ref{fig:unsharp1}) shows several
surface brightness depressions beyond the inner bubbles.  All are within
the surface brightness edge shown in Figure~\ref{fig:fullimg}.
Although these depressions show no correlation with the existing radio
data, some of them are more statistically significant than the
depressions associated with the bubbles, which are clearly real and
seen in radio observations.  We therefore make the assumption that
these features are ``ghost cavities'', leftover from previous AGN
outbursts, devoid of X-ray emitting gas but no longer strongly
emitting in the radio, as seen in
other systems (\eg, Jetha \etal\ 2008).  We estimate the total volume
occupied by the cavities by extracting spectra from the central
30\arcsec\ (22~kpc), with and without including the regions of the
cavities, and fitting with an absorbed APEC model.  Comparing the volume
emissivities (calculated from the area-normalized normalizations of the
fits) directly gives an estimate for the fraction of the total volume
occupied by the cavities.  We find a total cavity volume of 1240~kpc$^3$
for 14 cavities within the central 22~kpc (\ie, a filling-factor of
about 3\%), giving an average cavity volume of 31~kpc$^3$ (note that
not all of these cavities are visible in Figure~\ref{fig:unsharp1},
as some are within the large dark feature to the east that
corresponds to an edge in the central surface brightness, and are not
visible at this scaling and contrast).  For
spherical cavities, this gives an average radius of $\sim2$~kpc, fully
consistent with the average projected cavity radius (even though
several cavities are clearly non-spherical) and with our assumption
that the cavities contain no X-ray emitting gas.  Assuming that the
bubbles rise buoyantly at the sound speed in the gas, $c_s \approx
460\kms$ for $kT = 0.8$~keV, the average distance of the bubbles from
the AGN $d_{\rm avg} = 9$~kpc gives an average cavity age of $t_{\rm
  age} \approx 18$~Myr.

The mechanical energy input required to inflate the cavities, $E_{\rm
  mech}$, can be estimated from the X-ray gas pressure and the total
volume occupied by the cavities.  Using the density and
temperature from the spectral fit to the central 22~kpc, we find an
average pressure of $\sim1.8 \times 10^{-11}$~dyne~cm$^{-2}$, which
gives a total mechanical energy input of $E_{\rm mech} \approx 7
\times 10^{56}$~ergs,
consistent with what is found in other galaxy groups (McNamara 2004).
We obtain an estimate for the cooling rate of the central gas from the
0.3--12~keV X-ray luminosity 
in the same region, which we find from our fitted spectral model to be
$L_X = 1.5 \times 10^{42}$~ergs~s$^{-1}$. 
The mechanical energy in the cavities is therefore sufficient to offset cooling
in the diffuse gas for 15~Myr, very similar to what was found for the
Virgo galaxy NGC~4552 (Machacek \etal\ 2006), and similar to the
average cavity age calculated from the rise time above.  This is on the order of
central AGN cycle times inferred for other galaxy clusters (\eg,
Blanton \etal\ 2009; Clarke \etal\ 2009) of a few tens of Myrs.  We
therefore conclude
that the current average mechanical luminosity of the AGN
is nearly if not completely sufficient to balance radiative cooling of
the gas in the central region (we also note that the total energy in
the bubbles may be up to a factor of 2--4 times more than the
mechanical energy alone, see McNamara 2004).

\section{Summary} \label{sec:summary}

We have analyzed {\it Chandra} ACIS-S3 observations of the NGC~5098
galaxy group.  X-ray images reveal a spiral arm morphology extending to the
northeast, a sharp edge in the diffuse emission surrounding much of
NGC~5098a and connecting to the outer boundary of the arm, as well as
bubbles and other X-ray surface brightness depressions in 
the core.  Temperature and density profiles across the edges indicate
that they are cold fronts.  The structure of the cold fronts (which
have relatively modest density jumps, and radii of curvature centered
on the peak of the diffuse emission), the spiral structure seen in the
X-ray surface brightness and temperature maps, and the offset between
the central clump of cool gas seen in the temperature map from the
central AGN and optical center of the galaxy all point to gas
sloshing in the core, as seen in simulations and other observations.
The obvious candidate perturber is the nearby galaxy NGC~5098b, which
has apparently been stripped of its X-ray emitting gas.  We have
suggested that the long outer arm of cool gas seen in the temperature map
may in part be the stripped tail of NGC~5098b, formed as it passed
through the group ICM and initiated sloshing in NGC~5098a.  The
winding direction of the inner spiral arm is consistent with
the perturber approaching from the northeast to the southeast, which
matches the trajectory implied for NGC~5098b if we interpret the
extended temperature map arm as its stripped tail.  The
two bubbles in the core of NGC~5098a, which are seen as X-ray cavities
surrounded by bright rims of emission, correlate with radio
observations.  A comparison of the radio pressure in the bubbles to
the X-ray pressure in the rims shows that they are about equal,
consistent with these being young features that are currently being
inflated by jets from the central AGN.  We make the assumption that the
other X-ray cavities seen in the core are ghost cavities, left over
from previous AGN outbursts.  An estimate of the mechanical energy
required to inflate these cavities shows that the energy output of the
central AGN is sufficient to balance radiative cooling of the gas in
this region.

\acknowledgments
The financial support for this
work was partially provided for by the Chandra X-ray Center through
NASA contract NAS8-03060, and the Smithsonian Institution. We thank
the anonymous referee for useful comments.

\clearpage

\begin{deluxetable}{lcccc}
\tablewidth{7.0truein}
\tablecaption{Spectral Fits \label{tab:spectra}}
\tablehead{
\colhead{Region \#}&
\colhead{$kT$}&
\colhead{Abund.}&
\colhead{$\chi^2$/dof}&
\colhead{Net Cnts.}\\
\colhead{}&
\colhead{(keV)}&
\colhead{(solar)}&
\colhead{}&
\colhead{}
}
\startdata
R1&0.934$^{+0.014}_{-0.014}$&0.30$^{+0.04}_{-0.03}$&50.2/50=1.00&3743\\
R2&0.768$^{+0.045}_{-0.058}$&0.17$^{+0.11}_{-0.07}$&3.6/7=0.82&554\\
R3&0.813$^{+0.043}_{-0.051}$&0.20$^{+0.13}_{-0.08}$&2.8/7=0.90&484\\
R4&0.993$^{+0.030}_{-0.033}$&0.26$^{+0.11}_{-0.08}$&4.7/7=0.69&793\\
R5&1.223$^{+0.046}_{-0.054}$&0.58$^{+0.28}_{-0.18}$&6.55/7=0.48&687\\
R5&1.147$^{+0.056}_{-0.060}$&(0.26)&11.1/8=1.38&687\\
R6&1.056$^{+0.066}_{-0.042}$&0.13$^{+0.04}_{-0.03}$&27.8/32=0.68&791\\
R7&1.298$^{+0.062}_{-0.076}$&0.20$^{+0.06}_{-0.08}$&25.5/18=1.42&546\\
\enddata
\end{deluxetable}

\begin{references}

\reference{}
Adelman-McCarthy, J., \etal\ 2008, ApJS, 175, 297

\reference{}
Ascasibar, Y., \& Markevitch, M. 2006, ApJ, 650, 102

\reference{}
B\^{i}rzan, L., McNamara, B. R., Nulsen, P. E. J., Carilli, C. L., \&
Wise, M. W. 2008, ApJ, 686, 859

\reference{}
Blanton, E. L., Randall, S. W., Douglass, E. M., Sarazin, C. L.,
Clarke, T. E. \& McNamara, B. R. 2009, ApJL, submitted

\reference{}
Blanton, E. L., Sarazin, C. L., \& McNamara, B. 2003, ApJ, 585, 227

\reference{}
Buote, D. A. 2000, MNRAS, 311, 176

\reference{}
Clarke, T. E., Blanton, E. L., Sarazin, C. L., Anderson, L. D.,
Gopal-Krishna, Douglass, E. M., Kassim, N. E. 2009, arXiv:0904.1610

\reference{}
Colla, G.,  \etal\ 1970, A\&AS, 1, 281

\reference{}
Condon, J. J., Cotton, W. D., Greisen, E. W., Yin, Q. F., Perley, R. A., Taylor, G. B., \& Broderick, J. J. 1998, AJ, 115, 1693

\reference{}
Croston, J. H. \etal\ 2009, arXiv:0901.1346

\reference{}
Davis, D. S., Mulchaey, J. S., \& Mushotsky, R. F. 1999, ApJ, 511, 34

\reference{}
Dupke, R., White, R. E., \& Bregman, J. N. 2007, ApJ, 671, 181

\reference{}
Fabian, A. C., Sanders, J. S., Taylor, G. B., Allen, S. W., Crawford, C.
S., Johnstone, R. M., Iwasawa, K. 2006, MNRAS, 366, 417

\reference{}
Gastaldello, F., Buote, D. A., Humphrey, P. J., Zappacosta, L.,
Bullock, J. S., Brighenti, F., \& Mathews, W. G. 2007, ApJ, 669, 158

\reference{}
Gastaldello, F., Buote, D. A., Temi, P., Brighenti, F., Mathews,
W. G., \& Ettori, S. 2009, ApJ, 693, 43

\reference{}
Gitti, M., McNamara, B. R., Nulsen, P. E. J., Wise, M. W. 2007, ApJ, 660, 1118

\reference{}
Govoni, F. \& Feretti, L. 2004, IJMPD, 13, 1549

\reference{}
Hwang, U., Mushotzky, R. F., Burns, J. O., Fukazawa, Y., \& White,
R. A. 1999, ApJ, 516, 604 

\reference{}
Jetha, N. N., Hardcastle, M. J., Babul, A., O'Sullivan, E., Ponman,
T. J., Raychaudhury, S., Vrtilek, J. 2008, MNRAS, 384, 1344

\reference{}
Machacek, M., Nulsen, P. J. E., Jones, C., \& Forman, W. R. 2006, ApJ,
648, 947

\reference{}
Mahdavi, A., B\"{o}hringer, H., Geller, M. J., \& Ramella, M. 2000,
ApJ, 534, 114

\reference{}
Mahdavi, A., Finoguenov, A., B\"{o}hringer, H., Geller, M. J., \&
Henry, J. P. 2005, ApJ, 622, 187

\reference{}
Markevitch, M., Gonzalez, A., David, L., Vikhlinin, A., Murray, S.,
Forman, W., Jones, C., \& Tucker, W. 2002, ApJL, 567, 27

\reference{}
Markevitch, M., \& Vikhlinin, A. 2007, PhR, 443, 1

\reference{}
Mazzotta, P., Markevitch, M., Vikhlinin, A., Forman, W. R., David,
L. P., \& VanSpeybroeck, L. 2001, ApJ, 555, 205

\reference{}
McNamara, B. R. 2004, {\it Clusters of Galaxies: Probes of
  Cosmological Structure and Galaxy Evolution} (Carnegie
Observatories, Pasadena)

\reference{}
McNamara \& Nulsen, 2007, ARA\&A, 45, 117

\reference{}
Morganti, R., Parma, P., Capetti, A., Fanti, R., de Ruiter, H. R., \&
Prandoni, I. 1997, A\&A, 126, 335

\reference{}
Nulsen, P. J. E., Hambrick, D. C., McNamara, B. R., Rafferty, D.,
Birzan, L., Wise, M. W., \& David, L. P. 2005, ApJ, 625, L9

\reference{}
Pacholczyk, A. G. 1970, {\it Radio Astrophysics} (W. H. Freeman \&
Co., San Francisco)

\reference{}
Parma, P., de Ruiter, H. R., Fanti, C., Fanti, R. 1986, A\&AS, 64, 135

\reference{}
Pedlar, A., Ghataure, H. S., Davies, R. D., Harrison, B. A., Perley,
R., Crane, P. C., \& Unger, S. W. 1990, MNRAS, 246, 477

\reference{}
Peterson, J. R., \& Fabian, A.C. 2006, Physics Reports, 427, 1

\reference{}
Ramella, M., Geller, M. J., \& Huchra, J. P. 1989, ApJ, 344, 57

\reference{}
Ramella, M., Geller, M. J., Huchra, J. P., \& Thorstensen, J. R. 1995,
AJ 109, 1458

\reference{}
Randall, S., Jones, C., Kraft, R., Forman, W. R., \& O'Sullivan,
E. 2009, ApJ, 696, 1431

\reference{}
Randall, S., Nulsen, P., Forman, W., Jones, C., Machacek, M., Murray,
S., \& Maughan, B. 2008, ApJ, 688, 208 

\reference{}
Sambruna, R. M., Eracleous, M., \& Mushotzky, R. F. 1999, ApJ, 526, 60

\reference{}
Smith, J., \etal\ 2002, AJ, 123, 2121

\reference{}
Sun, M., Voit, G. M., Donahue, M., Jones, C., Forman, W., \&
Vikhlinin, A. 2009, ApJ, 693, 1142

\reference{}
Xue, Y.-J., B\"{o}hringer, H., \& Matsushita, K. 2004, A\&A, 420, 833

\reference{}
Zhao, J.-H., Sumi, D. M., Burns, J. O., \& Duric, N. 1993, ApJ, 416, 51

\end{references}
\end{document}